\documentclass[aps,pre]{revtex4}
\usepackage[dvips]{graphics}
\usepackage{graphicx}
\usepackage{amsfonts}
\usepackage{amssymb}
\usepackage{amsmath}
\usepackage{subfigure}
\usepackage{color}
\usepackage{times}
\usepackage{epsfig}

\begin{document}

\title{Quenched dynamics of two-dimensional solitons and vortices in the
Gross-Pitaevskii equation}
\author{Qian-Yong Chen}
\affiliation{Department of Mathematics and Statistics, University of Massachusetts,
Amherst MA 01003-4515, USA}
\author{P. G.\ Kevrekidis }
\affiliation{Department of Mathematics and Statistics, University of Massachusetts,
Amherst MA 01003-4515, USA}
\author{Boris A. Malomed}
\affiliation{Department of Physical Electronics, School of Electrical Engineering,
Faculty of Engineering, Tel Aviv University, Tel Aviv 69978, Israel}

\begin{abstract}
We consider a two-dimensional (2D) counterpart of the experiment
that led to the creation of quasi-1D bright solitons in
Bose-Einstein condensates (BECs) [Nature \textbf{417}, 150--153
(2002)]. We start by identifying the ground state of the 2D
Gross-Pitaevskii equation for repulsive interactions, with a
harmonic-oscillator (HO) trap, and with or without an optical
lattice (OL). Subsequently, we switch the sign of the interaction to
induce interatomic
attraction and monitor the ensuing dynamics. Regions of the stable
self-trapping and catastrophic collapse of 2D fundamental solitons
are identified in the parameter plane of the OL strength and BEC
norm. The increase of the OL strength expands the persistence domain
for the solitons to larger norms. For single-charged solitary
vortices, in addition to the survival and collapse regimes, an
intermediate one is identified, where the vortex resists the
collapse but loses its structure, transforming into a fundamental
soliton. The same setting may also be implemented in the context of
optical solitons and vortices, using photonic-crystal fibers.
\end{abstract}

\maketitle

\section{Introduction}

The last decade has brought about a very substantial amount 
of research efforts in the physics
of atomic Bose-Einstein condensates (BECs) \cite{book1,book2}. These studies
have revealed a wide array of interesting phenomena, not only thanks to the
precise control over experimental settings and the use of accurate and
relatively simple theoretical models, which is a unique peculiarity of this
area \cite{rmp,ourbook}, but also due to direct connections to other areas
of physics, including superfluidity, superconductivity, quantum and
nonlinear optics, and nonlinear wave theory.

One of the main topics for which these connections have been pursued is the
study of the nonlinear dynamics of matter waves in BECs. Diverse
experimental techniques have been used to produce a broad array of
matter-wave excitations. In particular, phase engineering has been used to
create vortices \cite{chap01:vort1,chap01:williams} and dark solitons \cite%
{chap01:denschl,chap01:dutton,chap01:dark1,chap01:dark,engels}. Stirring of
the BECs has led to the formation of vortices \cite%
{chap01:vort2,chap01:vort3} and vortex-lattices \cite%
{chap01:latt1,chap01:latt2,chap01:latt3}. The switch of the scattering
length, from positive (repulsive) to negative (attractive), via Feshbach
resonances, has been used to produce bright matter-wave solitons
and soliton trains \cite%
{chap01:bright1,chap01:bright2,chap01:bright3,chap01:njp2003b}. These modes
have been studied extensively, to the extent that numerous reviews (and even
books~\cite{ourbook}) are dedicated to bright solitons~\cite{kono,fatk},
dark solitons~\cite{djf} and vortices~\cite{fetter,us,fetter2}.

In this work, we aim to revisit a fundamental aspect associated with some of
the principal experiments used to produce bright solitons, especially those
carried out by the Rice group~\cite{chap01:bright1,chap01:njp2003b}.
 (Note that a similar method was used for the creation of
solitons in optical fibers in the pioneering works in that field \cite{fiber1,fiber2}.)
Precisely, we study the modulational instability~(MI)\ \cite{djf_rev} of 
the fundamental and vortex soliton after the \textit{%
quench}, i.e., after switching the BEC system from repulsive to attractive
interactions. While this mechanism was extremely efficient in the experimentally
elaborated cigar-shaped setting, to the best of our knowledge it has not yet
been systematically explored in quasi-two-dimensional (2D) pancake-shaped
BECs. This is the subject of the present work, with emphasis on the
formation of solitons and solitary vortices. More specifically, we examine
the results of the sign switching of  the nonlinearity from repulsive to
attractive in the 2D BEC, trapped via a combination of a
harmonic-oscillator (HO), i.e., magnetic, and periodic optical-lattice~(OL)\
potentials \cite{kono,morsch}. Our initial condition, prior to the quench,
represents either the ground state of the system in the
repulsive-interaction regime, or the dynamically stable excited state in the
form of a vortex~with topological charge $S=1$ \cite{fetter,us,fetter2}. We
note in passing that the vortex is stable in the case of the HO trap if its
pivot is collocated with a local maximum of the OL potential, yet
potentially unstable if is placed at a minimum \cite{vortjpb,law}.

The paper is structured as follows. In section II, we introduce the model
and some essential features of the numerical analysis. In section III, we
report conclusions produced by the simulations for the fundamental solitons.
A time-dependent variational approximation (VA) for this case is elaborated
in section IV. In section V, the case of the solitary
vortex is considered. A summary of the results and a discussion of
possibilities for subsequent work are presented in section VI.

\section{The model and computational approach}

\label{sec:2D}

For sufficiently cold and dilute atomic gases, where the mean-field
approximation is well-established, the BEC dynamics can be described
by the mean-field order parameter (wave function), $\Psi (\mathbf{r},t)$.
Assuming a strongly anisotropic trap, with the transverse ($x,y$) and
longitudinal ($z$) trapping frequencies chosen so that $\omega _{x}=\omega
_{y}\equiv \omega _{\perp }\ll \omega _{z}$, the trapped BEC acquires a
nearly planar, (``pancake") shape~\cite{book1,book2,ourbook}. 
This, in turn, permits one to factorize the wave
function, $\Psi =\Phi (z)\psi (x,y)$, where $\Phi (z)$ is the ground state
of the respective HO. Next, averaging the underlying 3D Gross-Pitaevskii
equation (GPE) equation along the longitudinal direction, $z$, leads to the
following reduced (2D) GPE for the transverse component of the wave function
(see also Refs.~\cite{book1,book2,ourbook}):
\begin{equation}
i\hbar \partial _{t}\psi =-\frac{\hbar ^{2}}{2m}\Delta \psi +g_{\mathrm{2D}%
}|\psi |^{2}\psi +V_{\mathrm{ext}}(x,y)\psi . \label{gpe}
\end{equation}%
Here, $\Delta \equiv \partial _{x}^{2}+\partial _{y}^{2}$ is the 2D
Laplacian, $m$ is the atomic mass, and $g_{\mathrm{2D}}=g_{\mathrm{3D}%
}/\left( \sqrt{2}\pi a_{z}\right) $ is an effective 2D coupling constant,
where $g_{\mathrm{3D}}\equiv 4\pi \hbar ^{2}a_{s}/m$ ($a_{s}$ is the
scattering length), and the longitudinal trapping length is $a_{z}=\sqrt{%
\hbar /m\omega _{z}}$. The potential $V_{\mathrm{ext}}(x,y)$ in GPE~(\ref{gpe})
is a combination of the HO component and a 2D square-shaped OL:%
\begin{eqnarray}
V_{\mathrm{ext}}(x,y) &=&\frac{1}{2}m\omega _{\perp }^{2}r^{2}-\varepsilon
\lbrack \cos ^{2}(kx)+\cos ^{2}(ky)]  \notag \\[1ex]
&\equiv &V_{\mathrm{HO}}(r)+V_{\mathrm{OL}}(x,y).  \label{extpot}
\end{eqnarray}%
Here, $r$ is the radial variable, and the OL is characterized by its depth $%
V_{0}$ and periodicity $a_{\mathrm{OL}}=\pi /k$. 
The wavenumber $k=\left( 2\pi
/\lambda \right) \sin \left( \theta /2\right) $ (i.e., $a_{\mathrm{OL}%
}\equiv \lambda /\left( 2\sin \left( \theta /2\right) \right) $), in turn, is controlled by the wavelength of the interfering beams that create the lattice and angle $\theta $ between them.

Measuring length in units of $a_{\mathrm{OL}}/\pi \equiv 1/k$, time in units
of $\omega _{\mathrm{OL}}\equiv \hbar /E_{\mathrm{OL}}$, and energy in units
of $E_{\mathrm{OL}}\equiv 2E_{\mathrm{rec}}=\hbar ^{2}/ma_{\mathrm{OL}}^{2}$
(where $E_{\mathrm{rec}}$ is the lattice recoil energy), GPE~(\ref{gpe}) is
cast into the following dimensionless form:
\begin{equation}
iu_{t}+\frac{1}{2}\left( u_{xx}+u_{yy}\right) +g|u|^{2}u-V_{\mathbf{x}}u=0,
\label{eq:NLSE2D}
\end{equation}%
with potential 
\begin{equation}
V_{\mathbf{x}}=\frac{1}{2}\Omega ^{2}r^{2}-\varepsilon \left[ \cos 2x+\cos 2y%
\right] .  \label{eqnpot}
\end{equation}%
In this normalized GPE, the wave function is rescaled as $\psi \rightarrow
\sqrt{|g_{\mathrm{2D}}|/E_{\mathrm{OL}}}\psi \exp \left[ i(V_{0}/E_{\mathrm{%
OL}})t\right] $, and the sign parameter is $g\equiv \mathrm{sign}(g_{\mathrm{%
OL}})=\pm 1$, with $g=-1$ and $g=+1$ corresponding respectively, to
repulsive and attractive interatomic interactions. Further, the lattice
depth in Eq.~(\ref{eqnpot}) is measured in units of $4E_{\mathrm{rec}}$,
while the normalized HO strength is $\Omega =a_{\mathrm{OL}}^{2}/a_{\perp
}^{2}\equiv \omega _{\perp }/\omega _{\mathrm{OL}}$, where $a_{\perp }=\sqrt{%
\hbar /m\omega _{\perp }}$ is the transverse trapping length.

We are interested in the dynamics of fundamental solitons and solitary
vortices when the nonlinearity is switched from defocusing ($g<0$) to
focusing ($g>0$). More precisely, we first fix $g=-1$ and solve the
imaginary-time GPE version \cite{imaginary} to find the respective steady
state for a given configuration, and then solve the GPE with $g=1$, using
that steady state as the initial condition. All the parameters stay the same
except that $g$ changes sign. In all the presented examples, $\Omega =0.1
$ is assumed in Eq. (\ref{eqnpot}), yet our findings 
should be relevant to a wide range of $\Omega $'s within the
pancake setting.

Another way (a faster one) of finding such steady state that we employed
is to plug ansatz $u=e^{-i\mu t}v(x,y)$ into Eq.~(\ref%
{eq:NLSE2D}) to derive a nonlinear eigenvalue problem, which is then solved
with the Newton's method. Solutions produced by these two
approaches are found to be virtually identical, with the maximum pointwise difference 
being between $10^{-8}$ and $%
10^{-10}$ for most configurations.

Unless specified otherwise, the Newton's solution is used as
the initial condition in all the simulations. Further, the fourth-order split-step Fourier method~%
\cite{Blanes02, Montesinos05} is used to solve the GPE in time. The
corresponding domain size is $[-8\pi ,8\pi ]\times \lbrack -8\pi ,8\pi ]$,
with $256$ Fourier modes in each direction and time step $\Delta t=0.001$.
The simulations were run up to $T=2000$, which is large enough to
observe the stability of the final states (if they are stable). Figure~\ref%
{f:IC_S0S1} shows some examples of the initial conditions, i.e., the steady
states obtained by the Newton's method.

\begin{figure}[tbp]
\centerline{
    \psfig{file = 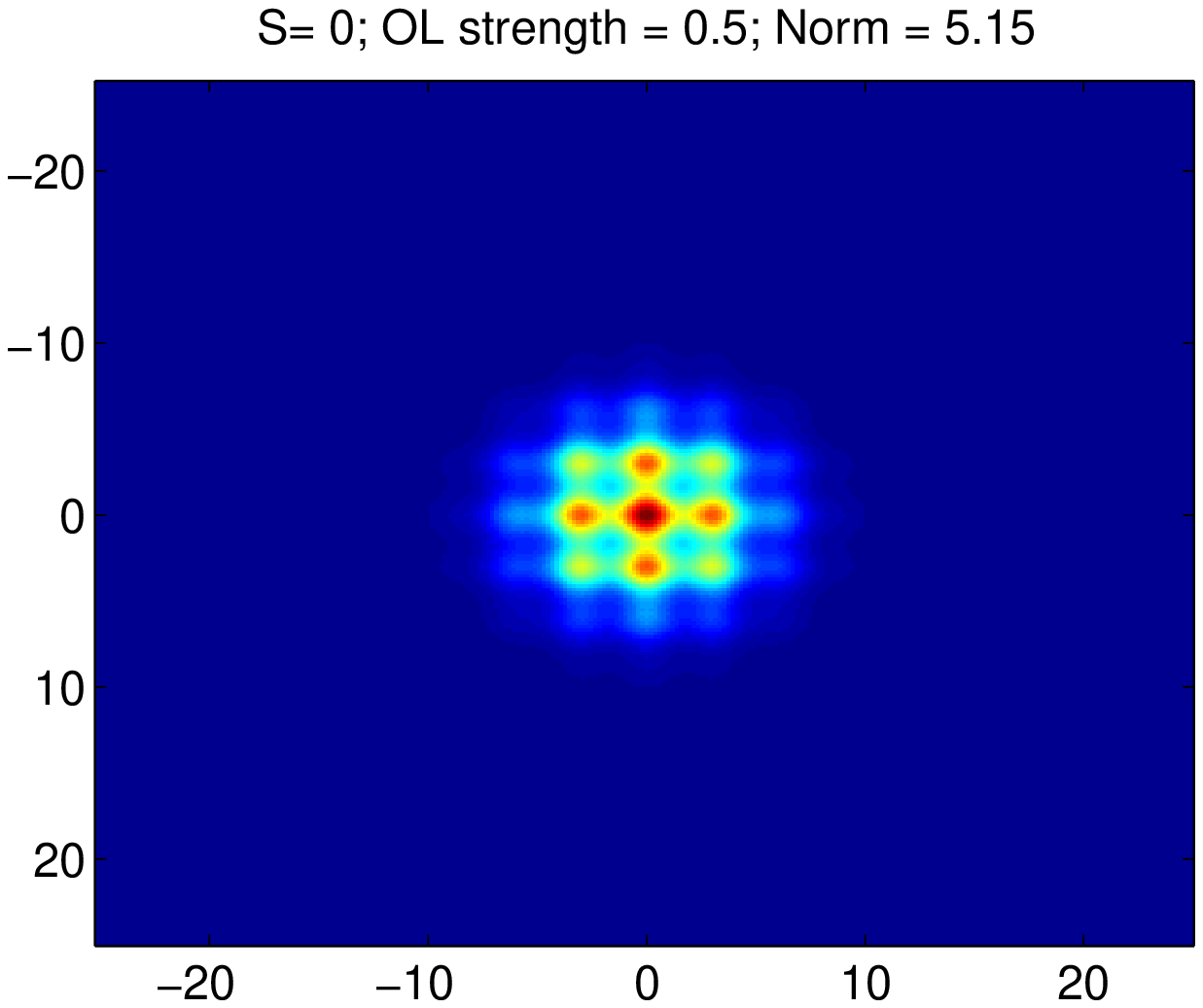, width=2.5in,height=2.5in}
\hspace{0.1in}
    \psfig{file = 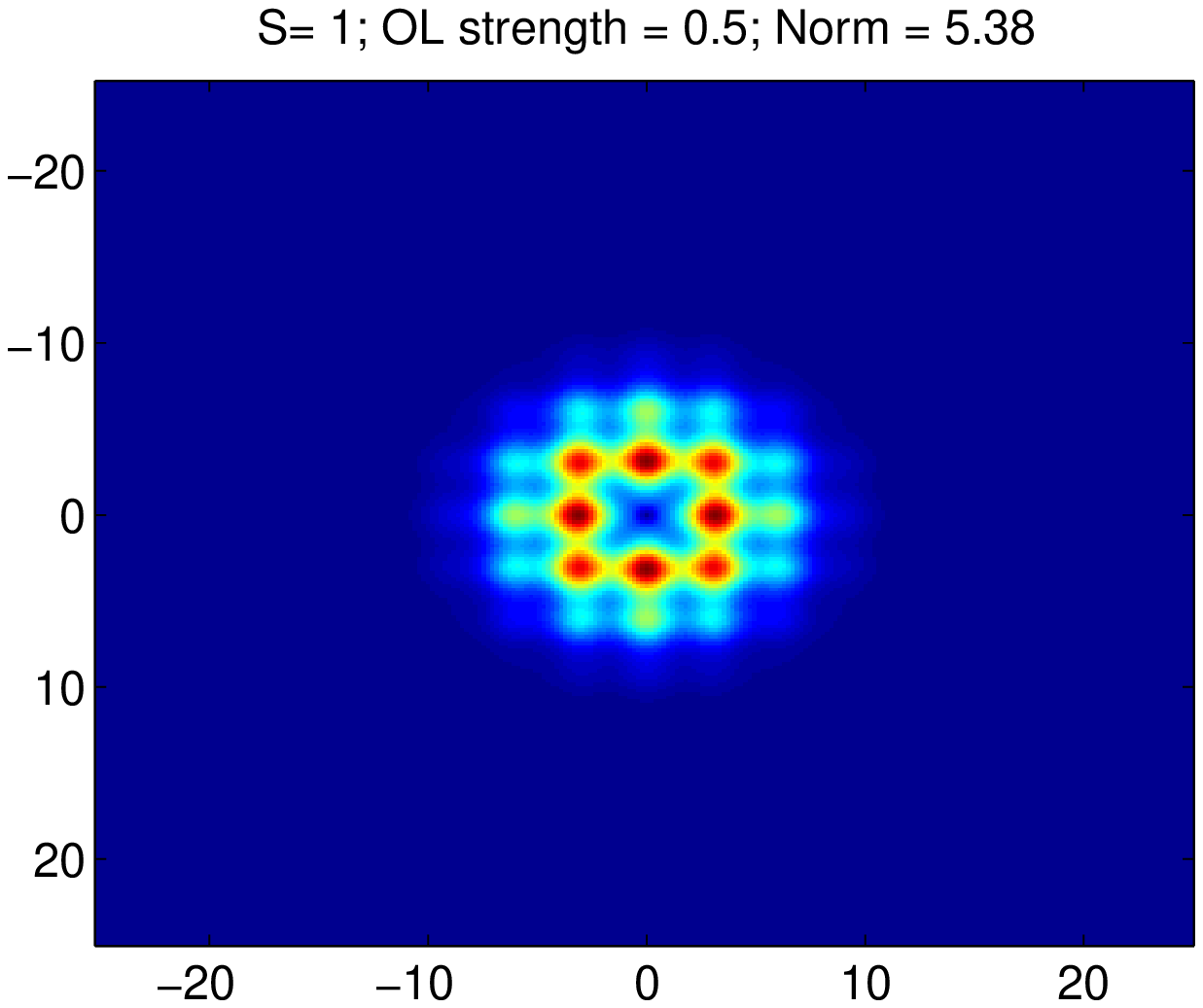, width=2.5in,height=2.5in}
}
\centerline{
    \psfig{file = 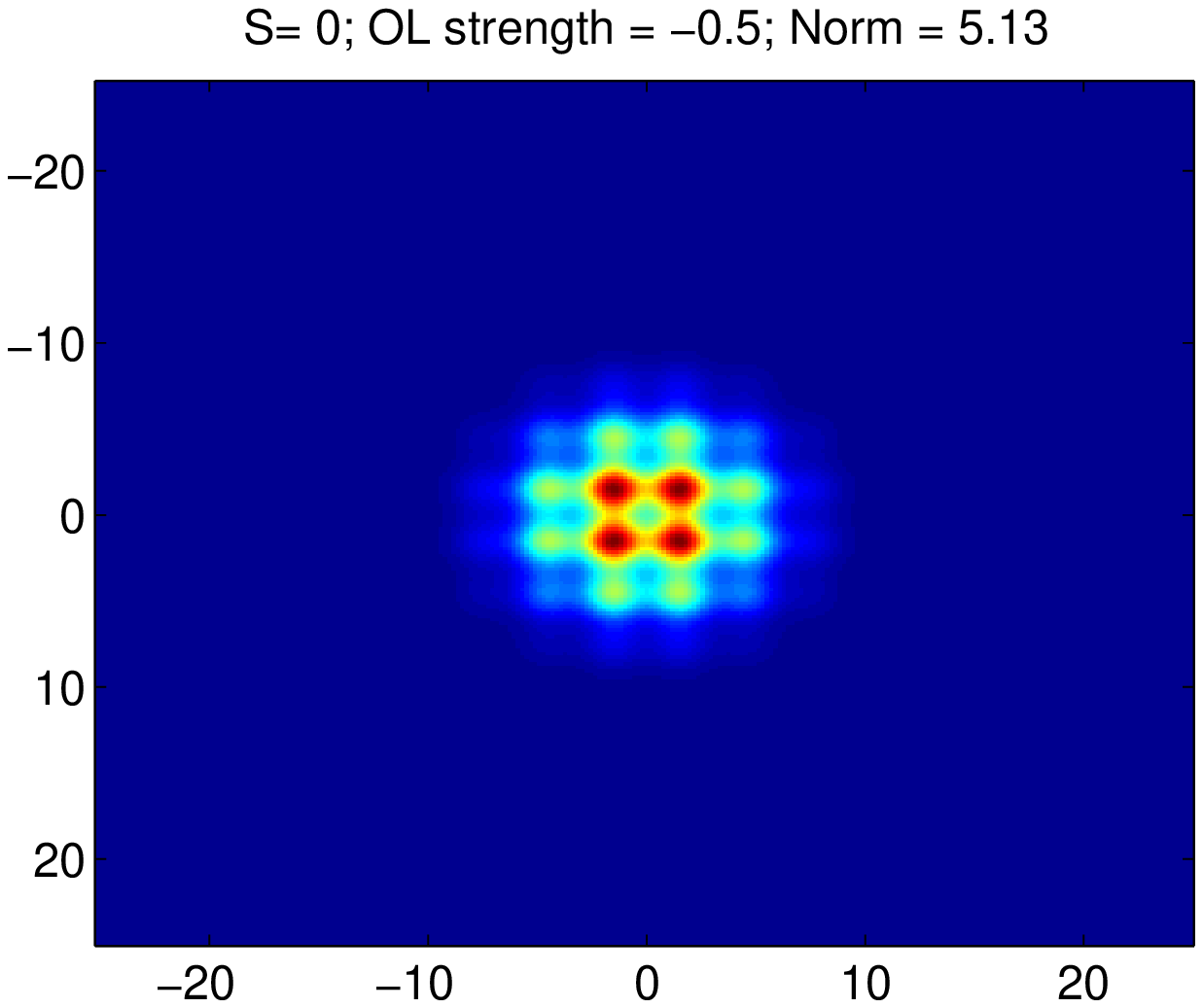, width=2.5in,height=2.5in}
\hspace{0.1in}
    \psfig{file = 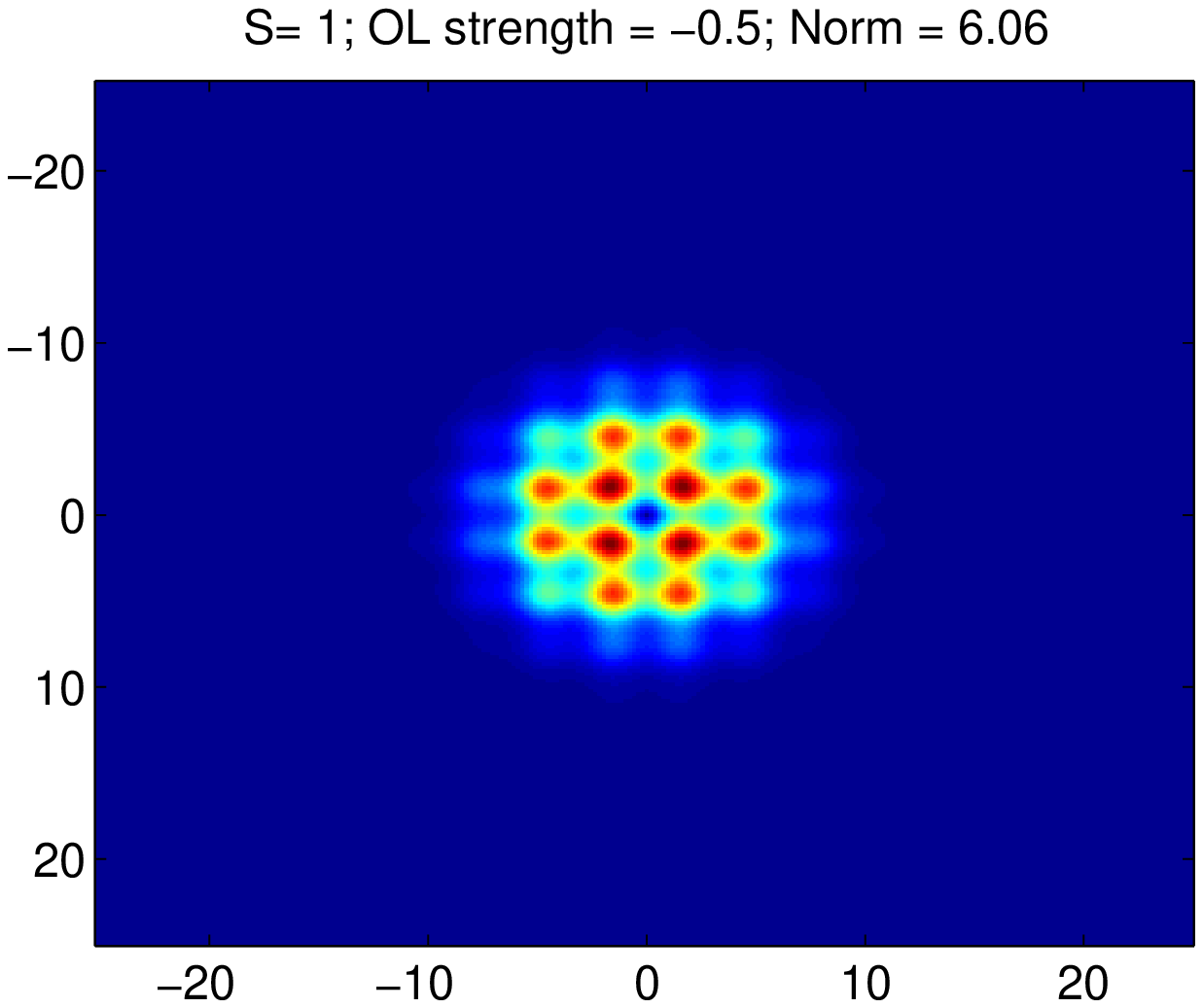, width=2.5in,height=2.5in}
 }
\caption{(Color online) A typical example of the output produced by
the Newton's method. Such states will be used as initial conditions
in sections III and V. Left panels: $S=0$ (fundamental solitons);
right panels: $S=1$ (vortices). Top and bottom panels correspond to
$\protect\varepsilon =0.5$ and $\protect\varepsilon =-0.5$,
respectively. } \label{f:IC_S0S1}
\end{figure}

Control parameters will be  the OL strength $\varepsilon $ and the norm of the initial condition 
(i.e., the normalized number of atoms)
\begin{equation}
N=\int \int \left\vert u_{0}\left( x,y\right) \right\vert ^{2}dxdy.
\label{N0}
\end{equation}%
The following main features of the
solution will be computed: amplitude $\left( |u|\right) _{\max }$, 
the final norm, which may be slightly different from the initial one due to
radiation losses, and the total angular momentum,
\begin{equation}
M= (-i) \int \int \left( \frac{\partial u}{\partial \theta }u^{\ast } -
 \frac{\partial u^{\ast}}{\partial \theta} u \right ) dxdy,  \label{M}
\end{equation}%
where $\theta $ is the angular coordinate. Note that $M$
is conserved in the isotropic system that does not include the
OL, and is not conserved with the presence of the OL. In particular, for
isotropic solutions with integer vorticity $S$,
\begin{equation}
u=e^{-i\mu t+iS\theta }U(r),
\end{equation}%
the relation between the angular momentum and norm is $M= 2 SN$. 
We consider both the fundamental soliton with $S=0$ and vortex soliton
with topological charge $S=1$, with or without the OL.

\section{Ground-state quench dynamics}

\begin{figure}[tbp]
\centerline{
    \psfig{file = 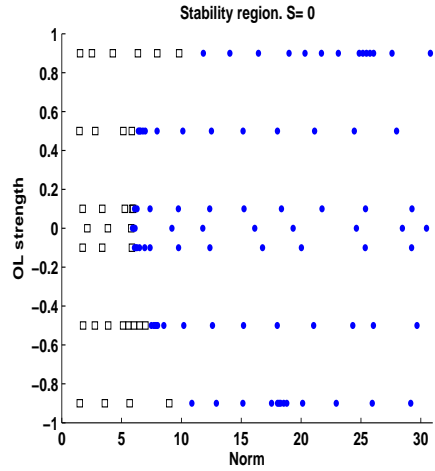, width=2.5in,height=2.5in}
   }
\caption{(Color online) The stability diagram for the fundamental soliton
resulting from the quench of the ground state of the repulsive BEC. The
squares denote stable configurations that support breathing dynamics, while
the dots denote unstable configurations that lead to the collapse.}
\label{f:SR_S0}
\end{figure}

We start with the fundamental state with no topological charge. We
perform the nonlinearity-sign switch (quench) for a wide range of OL
strengths and initial norms. The former is used as a representative parameter
associated with the potential, while the latter is employed for scanning
through the set of initial data. The resulting two-dimensional map of the
stability region of the ground state is plotted in Fig.~\ref{f:SR_S0}. 
There are two regions with one denoted by open squares, wherein the
ground state persists in the attractive regime in the form of a stable
bright 2D soliton (see, e.g., Refs.~\cite{carr2,uscarr} for detailed
discussions of the stability of such states), and the other denoted by
filled circles, which corresponds to the catastrophic wave collapse, a
well-known phenomenon for equations of the nonlinear-Schr{\"{o}}dinger type~%
\cite{sulem,Berge98}. Clearly, the OL plays a critical role towards the
stabilization the fundamental soliton, since the critical norm increases as
the OL gets stronger. When the OL is absent ($\varepsilon =0$), the soliton
exhibits breathing dynamics when $N\leq 5.81$, and will collapse at $N>5.91$.
Note that the collapse threshold corresponding to the Townes soliton is $N_{%
\mathrm{cr}}=5.85$ \cite{Berge98}.

\begin{figure}[tbp]
\centerline{
    \psfig{file = 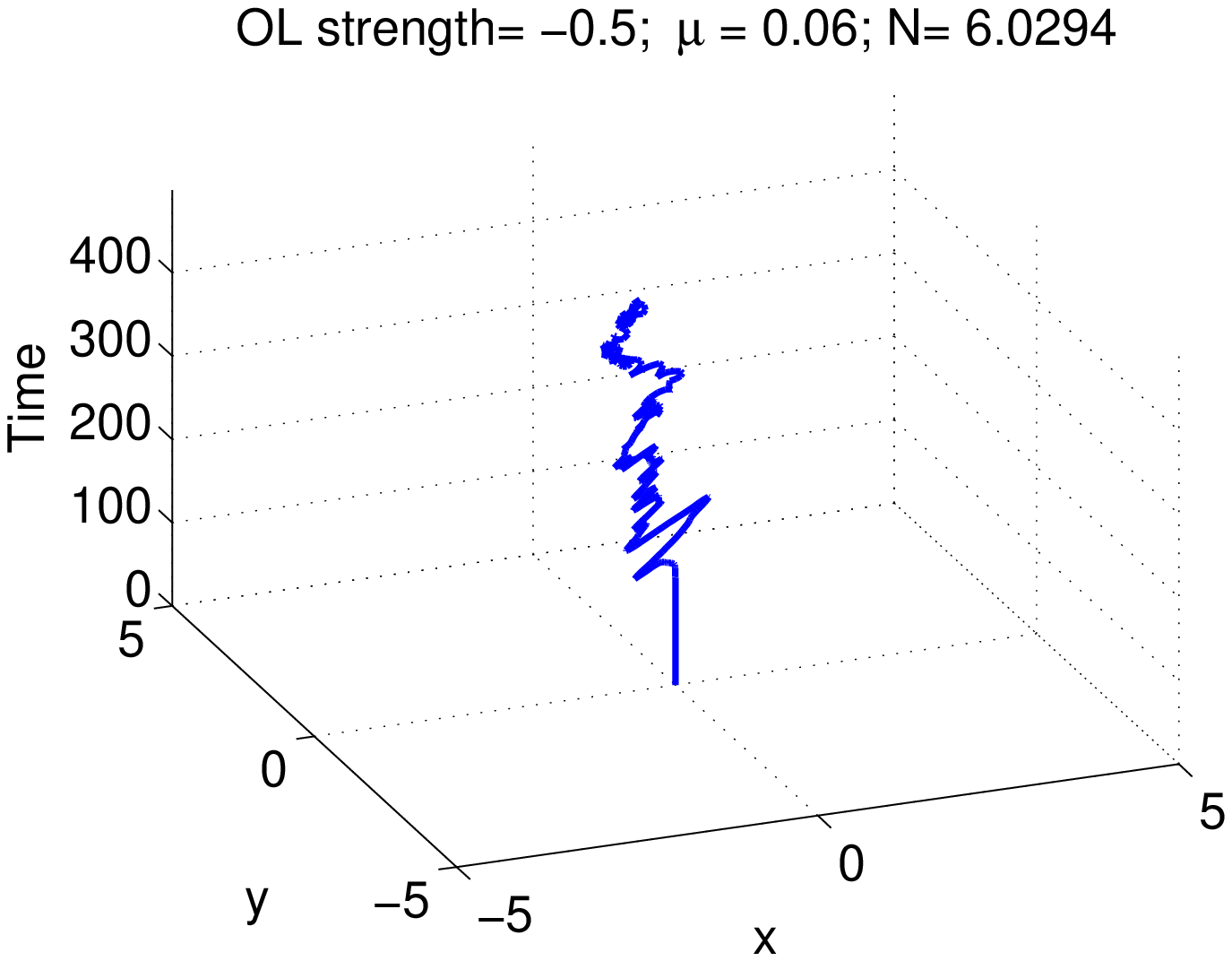, width=2.5in,height=2.5in}
\hspace{0.1in}
\psfig{file= 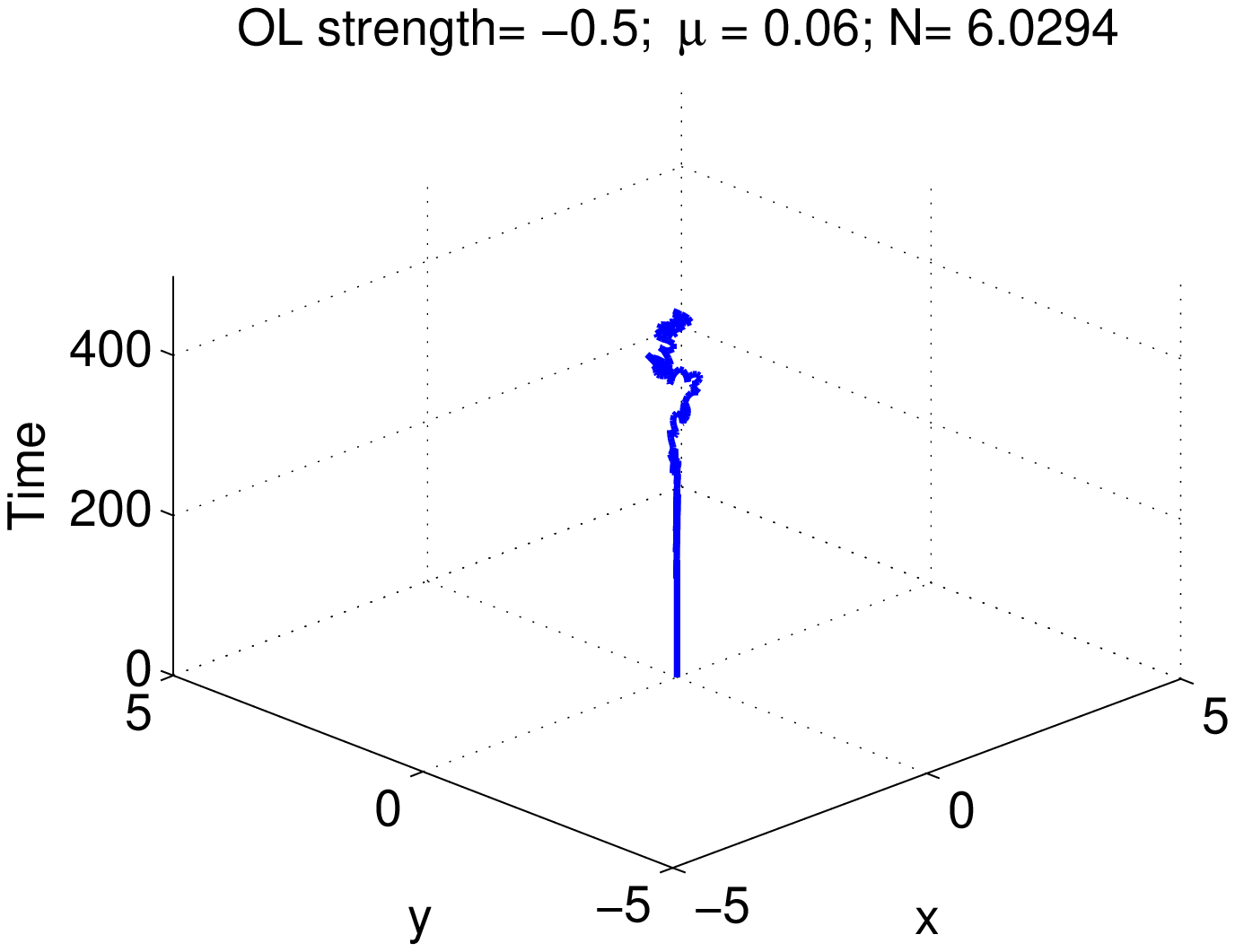, width = 2.5in, height = 2.5in}
}
\caption{(Color online) The trajectory of the soliton's center. The OL
strength is $\protect\varepsilon =-0.5$, while the soliton's parameters are $%
\protect\mu =0.06,~N=6.0294$. The left and right panels are shown from
different angles to better represent the rather complex motion.}
\label{f:massCenter}
\end{figure}

In the presence of the OL, the dynamics is more complex, and in
particular, the angular momentum will be generated when $\varepsilon <0$. 
In such cases, the soliton's center initially coincides with a local maximum
of the OL potential, hence it will slide down from this position. In fact,
in the beginning of the simulation, the soliton moves back and forth between
its initial position and the centers of other OL cells that are located along a
straight line. (It is relevant to note that a 2D soliton can travel more than
one cell, especially when the lattice is not very strong \cite{HS}.) 
Then the soliton starts to deviate from this straight line at certain time,
and the generation of the angular momentum commences.
The subsequent motion does not follow any simple pattern. The trajectory of
the soliton's center for one such case is shown in Fig.~\ref{f:massCenter}.

\section{The Variational Approximation}
\label{sec:VA}

To develop an understanding of the breathing regime exhibited by the
simulations, it is reasonable to apply the VA \cite{Progress}. The starting
point is the nonstationary GPE ({\ref{eq:NLSE2D}}), but with a
time-dependent nonlinearity\ coefficient $g(t)$.
This equation can be derived from the Lagrangian, $L=\int_{-\infty
}^{+\infty }dx\int_{-\infty }^{+\infty }dy\mathcal{L}(u)$, with density
\begin{gather}
\mathcal{L}=\frac{i}{2}\left( \frac{\partial u}{\partial t}u^{\ast }-\frac{%
\partial u^{\ast }}{\partial t}u\right) -\frac{1}{2}\left( \left\vert \frac{%
\partial u}{\partial x}\right\vert ^{2}+\left\vert \frac{\partial u}{%
\partial y}\right\vert ^{2}\right)   \notag \\
+\frac{g(t)}{2}|u|^{4}+\left\{ -\frac{1}{2}\Omega ^{2}r^{2}+\varepsilon %
\left[ \cos (2x)+\cos (2y)\right] \right\} |u|^{2}.  \label{density}
\end{gather}%
The variational \textit{ansatz} for the fundamental state is based on the
isotropic Gaussian,
\begin{equation}
u(r,t)=A(t)\exp \left( -\frac{r^{2}}{2W^{2}(t)}+\frac{1}{2}%
ib(t)\,r^{2}+i\phi (t)\right) ,  \label{ansatz}
\end{equation}%
where $A,W,b$ and $\phi $ are, respectively, the amplitude, width, chirp and
overall phase, which are assumed to be real functions of time. Following the
standard procedure, we insert the ansatz into density (\ref{density}) and
calculate the corresponding effective Lagrangian,
\begin{equation}
L_{\mathrm{eff}}=2\pi \int_{0}^{\infty }\mathcal{L}\,rdr.  \label{effective}
\end{equation}%
The result of the calculation is
\begin{eqnarray}
L_{\mathrm{eff}} &=&-N\frac{d\phi }{dt}-\frac{1}{2}NW^{2}\frac{db}{dt}-\frac{%
N}{2W^{2}}-\frac{1}{2}NW^{2}b^{2}  \notag \\
&&+\frac{N^{2}}{4\pi }\frac{g(t)}{W^{2}}-\frac{1}{2}\Omega
^{2}NW^{2}+2\epsilon Ne^{-W^{2}},  \label{L}
\end{eqnarray}%
where the overdot stands for the time derivative, and the norm of the ansatz
is, cf. Eq. (\ref{N0}):
\begin{equation}
N\equiv \int_{-\infty }^{+\infty }dx\int_{-\infty }^{+\infty }dy|u\left(
x,y\right) |^{2}=\pi A^{2}W^{2}.  \label{N}
\end{equation}%
The variational equation $\delta L_{\mathrm{eff}}/\delta \phi =0$ reproduces
the conservation of the norm, $dN/dt=0$, hence $N$ may be treated as a
constant. Then, equation $\delta L_{\mathrm{eff}}/\delta b=0$ yields an
expression for the chirp:
\begin{equation}
b=\frac{1}{W}\frac{dW}{dt},  \label{b}
\end{equation}%
and the final equation, $\delta L_{\mathrm{eff}}/\delta \left( W^{2}\right)
=0$, leads to a closed-form evolution equation for the width, in which $b$
is eliminated by means of Eq. (\ref{b}):
\begin{equation}
\frac{d^{2}W}{dt^{2}}=\frac{2\pi -Ng(t)}{2\pi W^{3}}\,-\Omega
^{2}W-4\epsilon We^{-W^{2}}.  \label{a2}
\end{equation}%
For constant $g>0$, Eq. (\ref{a2}) yields the well-known variational
approximation for the critical norm, $N_{\mathrm{cr}}^{\mathrm{(var)}}=2\pi
/g$~\cite{And}.

Aiming to compare results of the VA to those of the full simulations, we
have to note that, as the initial profiles in the direct simulations were
taken as per the steady state in the case with repulsive interactions, they
are much wider than the OL cell. For this reason, the model including the OL
cannot be adequately tackled by means of the (radially symmetric) ansatz (%
\ref{ansatz}), which does not include the density modulation induced by the
OL. Therefore, the comparison is only carried out for the fundamental state
in the absence of the OL.

\begin{figure}[tbp]
\centerline{
    \psfig{file = 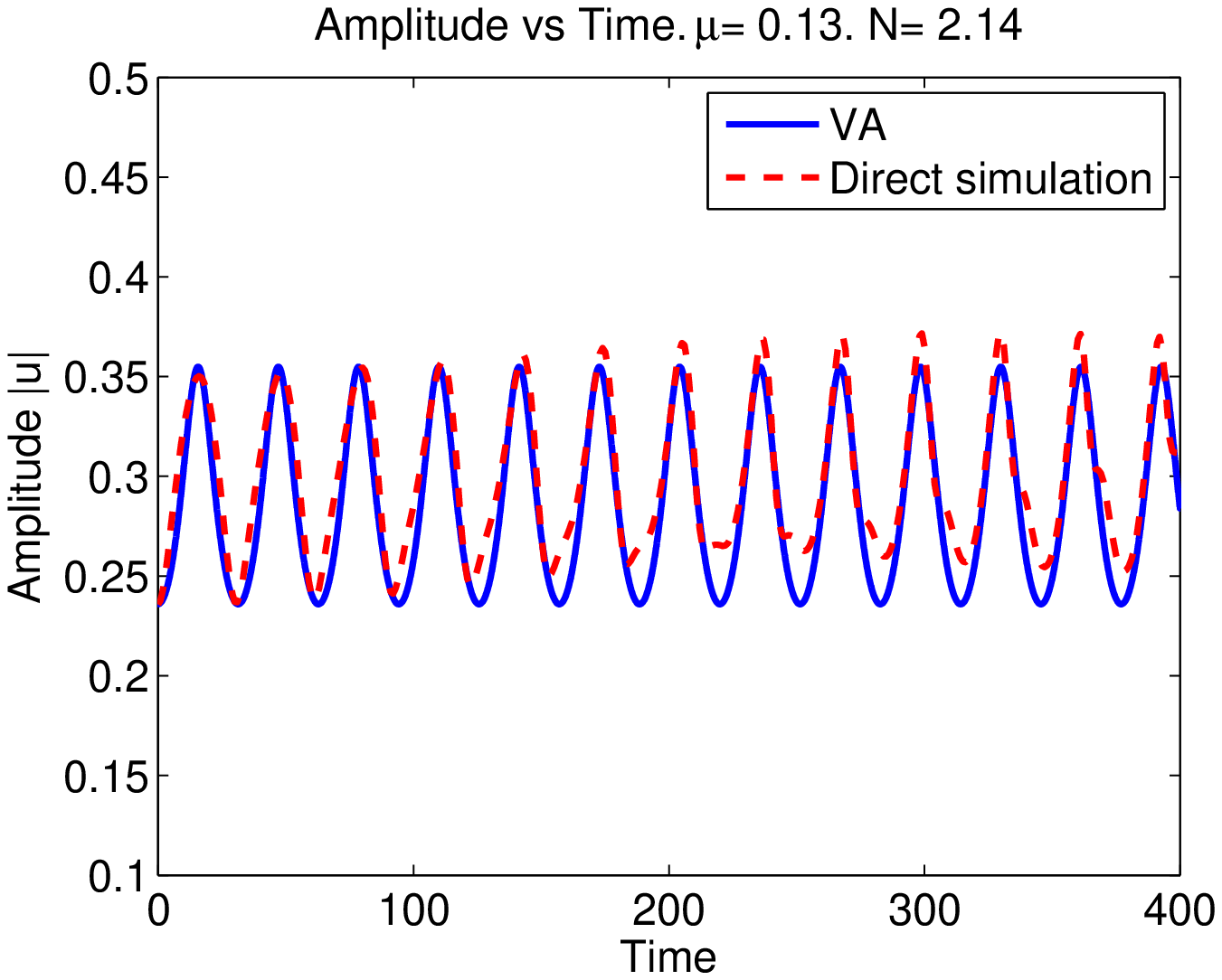, width=2.5in,height=2.5in}
\hspace{0.1in}
    \psfig{file = 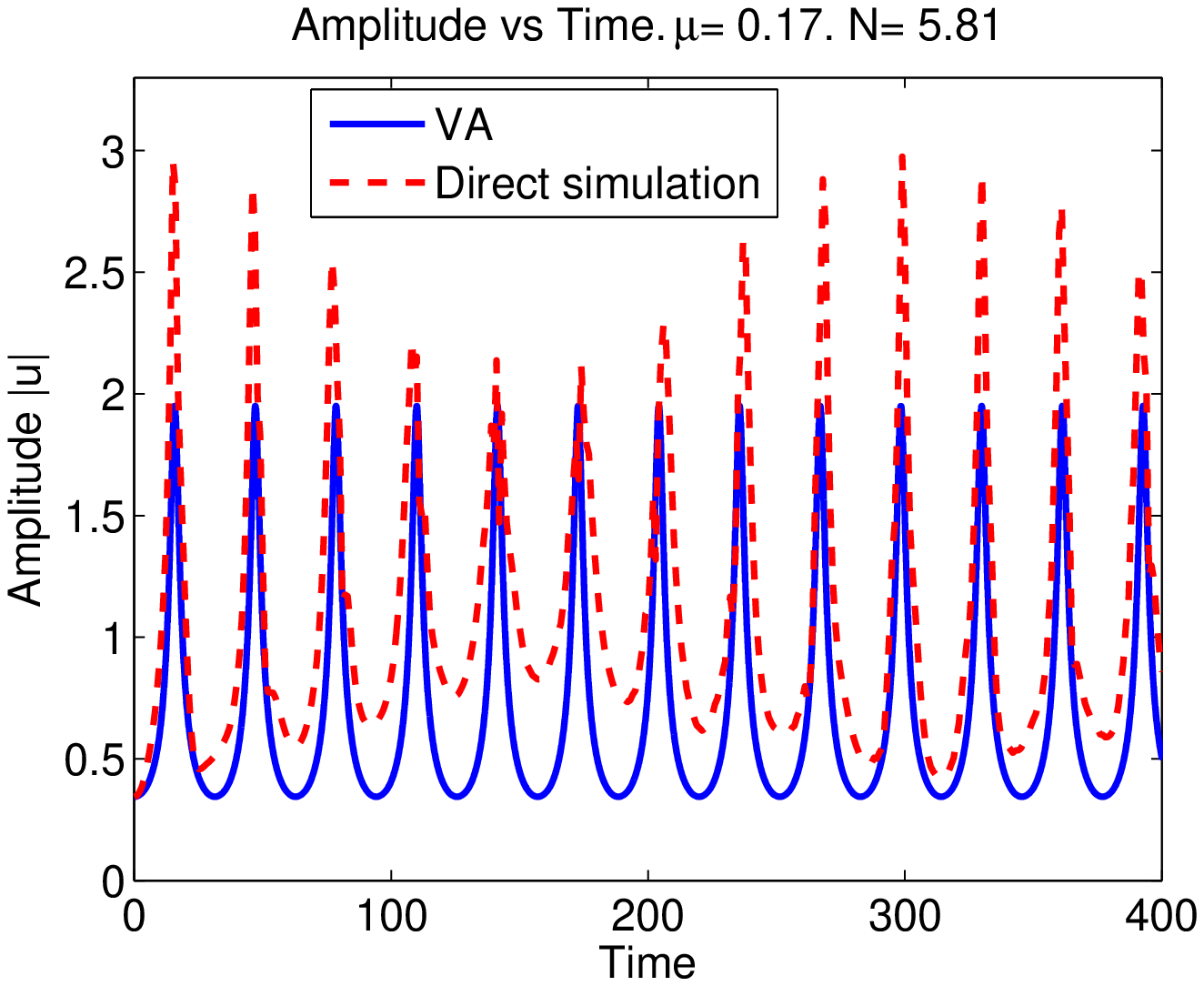, width=2.5in,height=2.5in}
   }
\caption{(Color online) Comparison of the amplitude of the breathing
fundamental soliton, generated by the quench, between the VA and direct
simulations. The left panel refers to $N=2.14$, and the right one to a
vicinity of the critical point, for $N=5.81$.}
\label{f:VA}
\end{figure}

A number of examples of the breathing regime of the resulting fundamental
solitons are displayed in Fig.~\ref{f:VA}. These are obtained by solving
Eq.~(\ref{a2}), for which two initial conditions are needed. 
The first initial condition $W(t=0)$ is simply the width of the initial soliton,
while the second initial condition $\frac{dW}{dt}(t=0)$
is approximated by a first order finite difference based on the widths of the solution
in direction simulation at $t=0$ and $t=0.001$.

In Fig.~\ref{f:VA}, one observes that, for sufficiently small $N$, the
resulting amplitude of the solution closely follows the VA\ prediction.
However, as $N$ increases, there arises a beating effect in the full GPE
dynamics that is presumably not accounted for by the simplified VA dynamics
of Eq.~(\ref{a2}). Nevertheless, the VA still captures the principal
features of the oscillatory dynamics of the width of the fundamental soliton.

\section{The quench-induced dynamics of vortices}

\begin{figure}[tbp]
\centerline{
    \psfig{file = 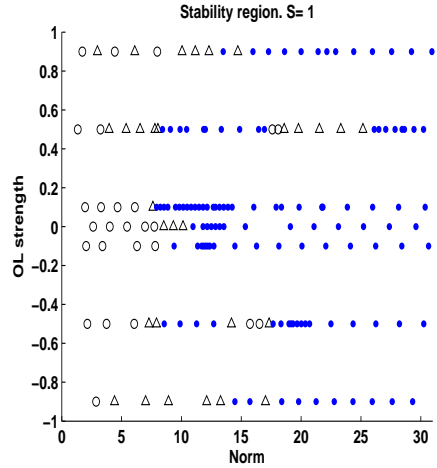, width=2.5in,height=2.5in}
   }
\caption{(Color online) The stability region for the vortical
initial condition with $S=1$. Here, three regimes are identified.
Circles represent the case when the resulting state is a vortex with
charge $S=1$. Triangles correspond to an intermediate regime of
non-collapsing solutions, which, however, do not keep the vorticity.
Finally, dots represent collapsing solutions.} \label{f:SR_S1}
\end{figure}

With the introduction of initial vorticity,
the most interesting observation is the rather delicate character of the stability of trapped solitary
vortices under the self-attractive nonlinearity (see Ref. \cite%
{Boris06} and references therein). For the vortical initial inputs, we have
performed an extensive numerical analysis similar to that reported in
section III for the simpler case of the fundamental state. The respective
two-parameter (the OL strength and initial norm) stability region is shown
in Fig.~\ref{f:SR_S1}. Here, we categorize both \textit{coherent} and
\textit{less coherent} (see the examples below) states generated by the
vortical inputs as stable when the collapse does not occur by $T=2000$ (our
\textit{reporting horizon}). When $\varepsilon =0$ (i.e., the OL is absent),
the critical norm for the quench process is found to be
\begin{equation}
N_{\mathrm{cr}}^{(S=1)}\approx 11.81,  \label{cr}
\end{equation}%
which is \emph{essentially larger} than the known value,$~N_{\max
}^{(S=1)}=7.79$, i.e., the boundary of the existence of 
(numerically) exact stable trapped
vortices with topological charge $S=1$ \cite{Boris06}. Thus, the
effective stability range of dynamical (breathing) vortices may be
essentially broader than that of their static counterparts. For the
simulation time exceeding $T=2000$, the critical norm
may be found to vary slightly, as the stable solutions found at $N$
very close to $N_{\mathrm{cr}}^{(S=1)}$ are still in the process of
splittings and recombinations at $T=2000$. These "hesitating"
solutions do not constitute a regular vortex soliton, but rather
exhibit an additional breathing process.

\begin{figure}[tbp]
\centerline{
    \psfig{file = 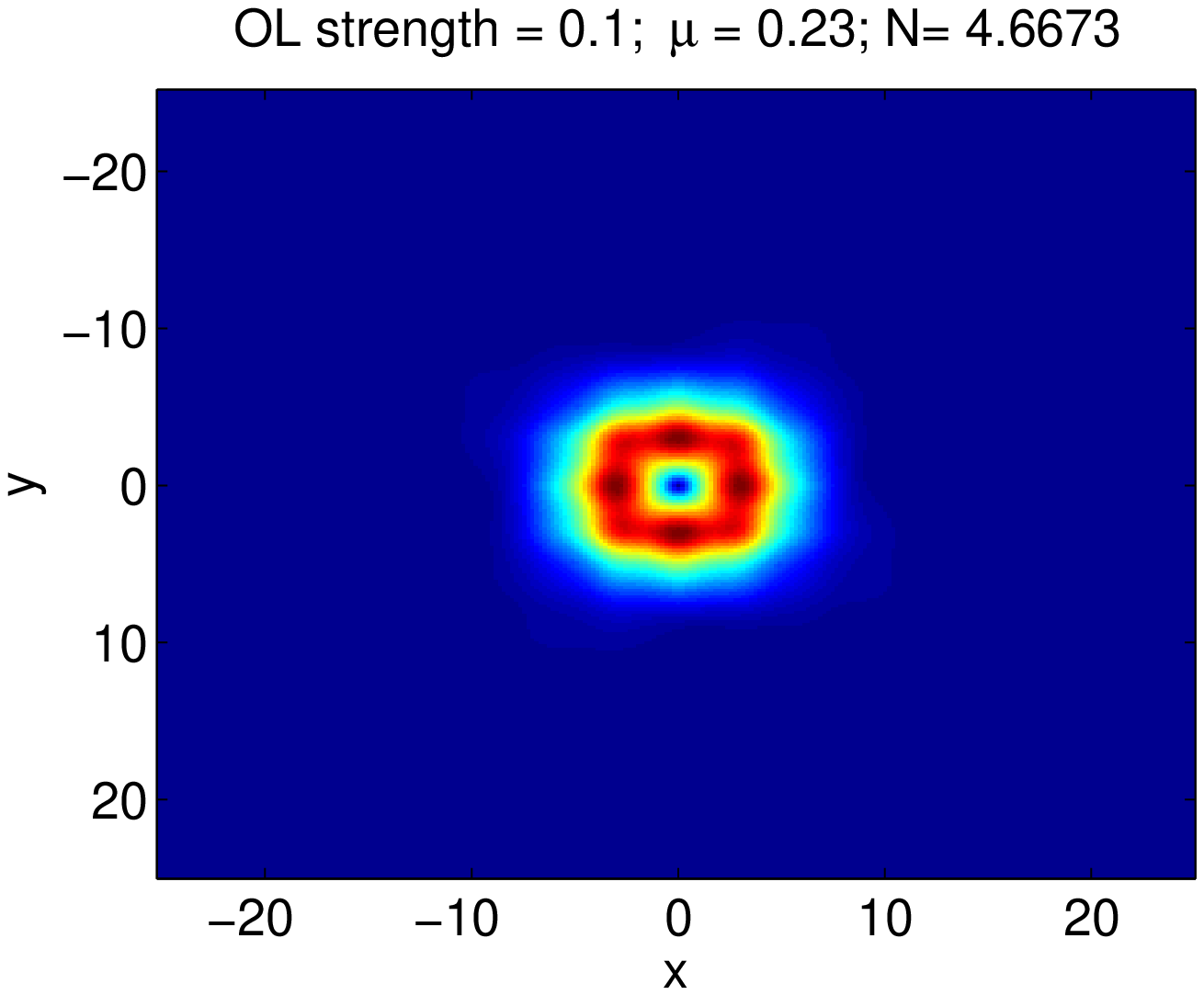, width=2.5in,height=2.5in}
\hspace{0.1in}
    \psfig{file = 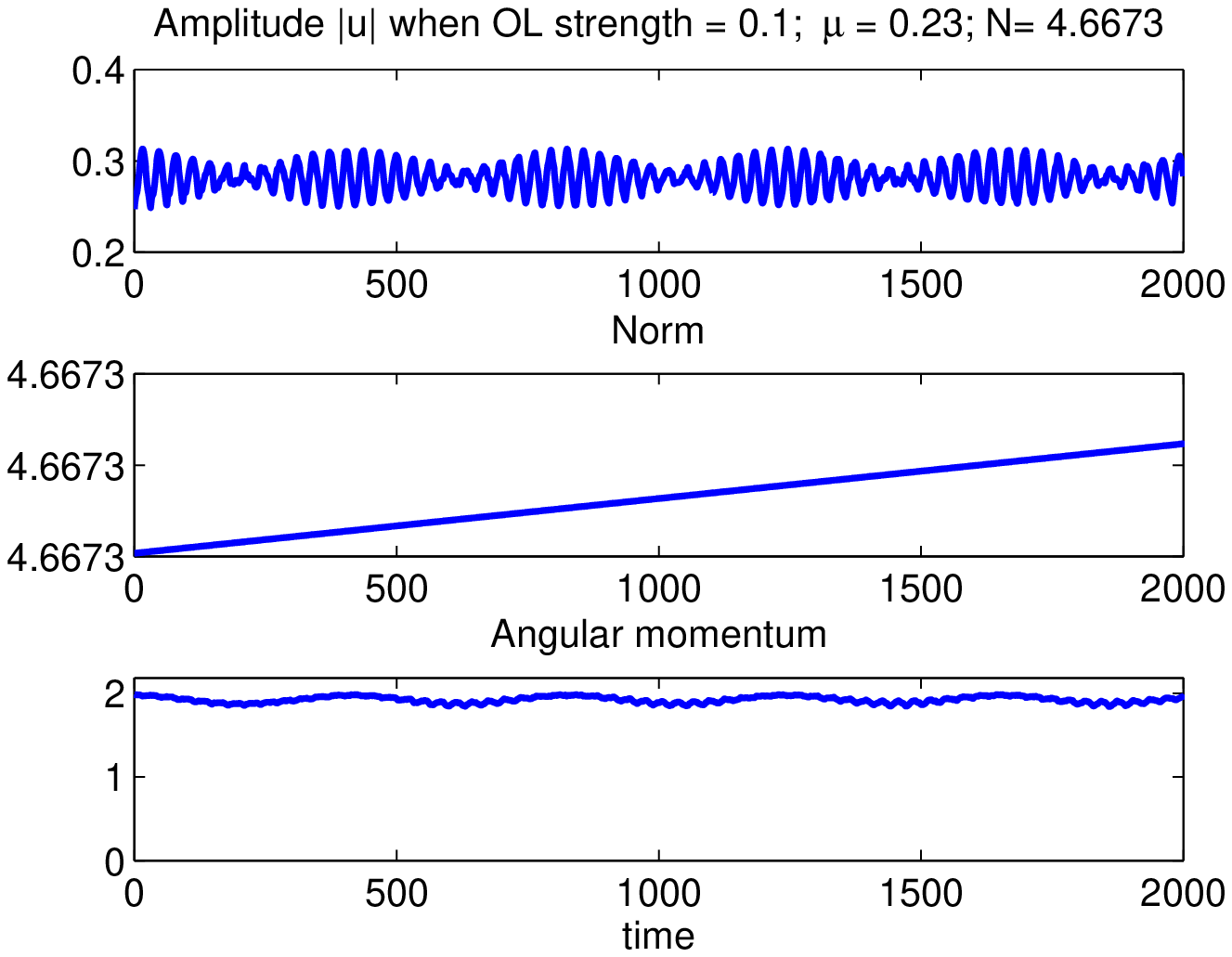, width=2.5in,height=2.5in}
   }
\centerline{
    \psfig{file = 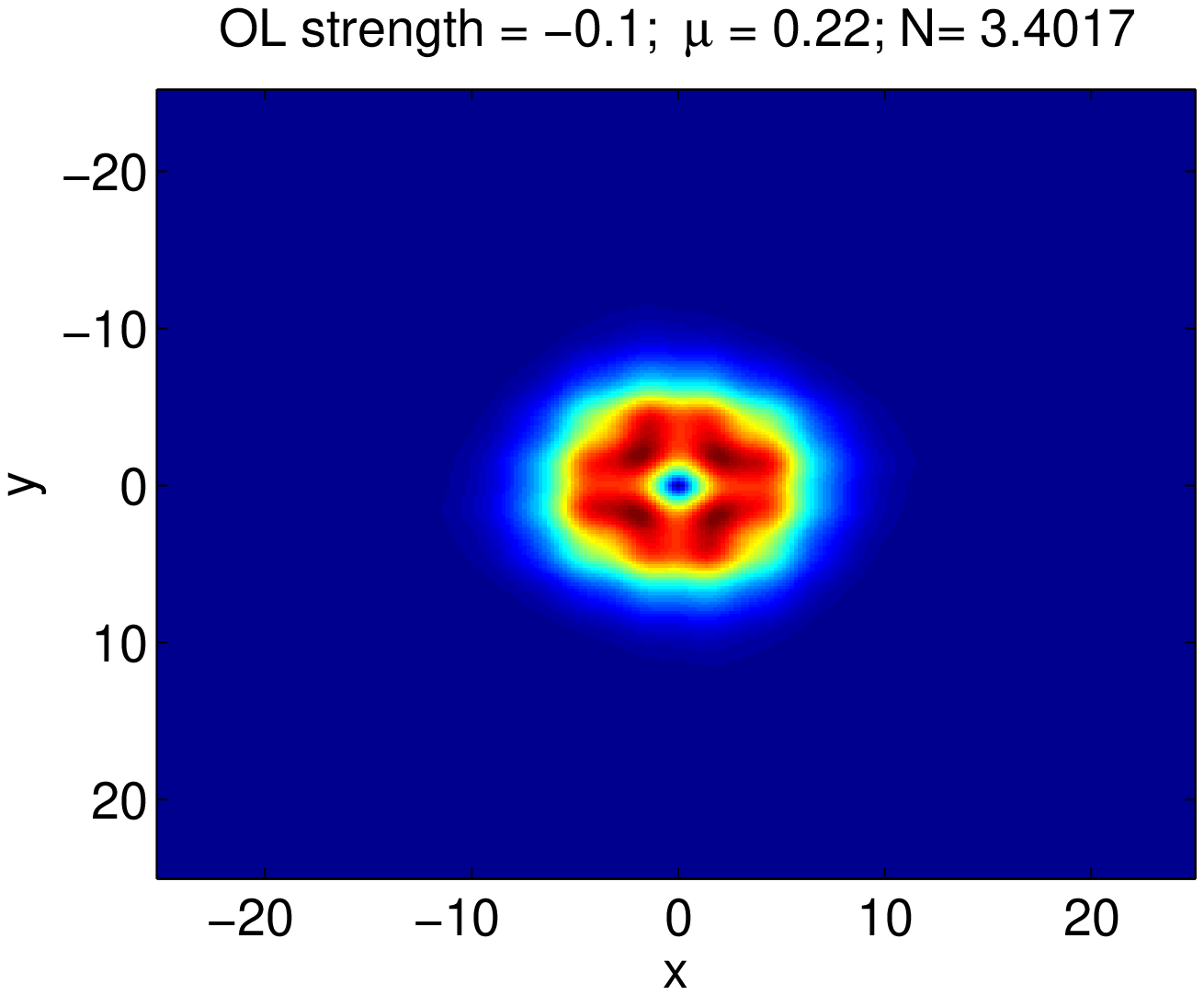, width=2.5in,height=2.5in}
\hspace{0.1in}
    \psfig{file = 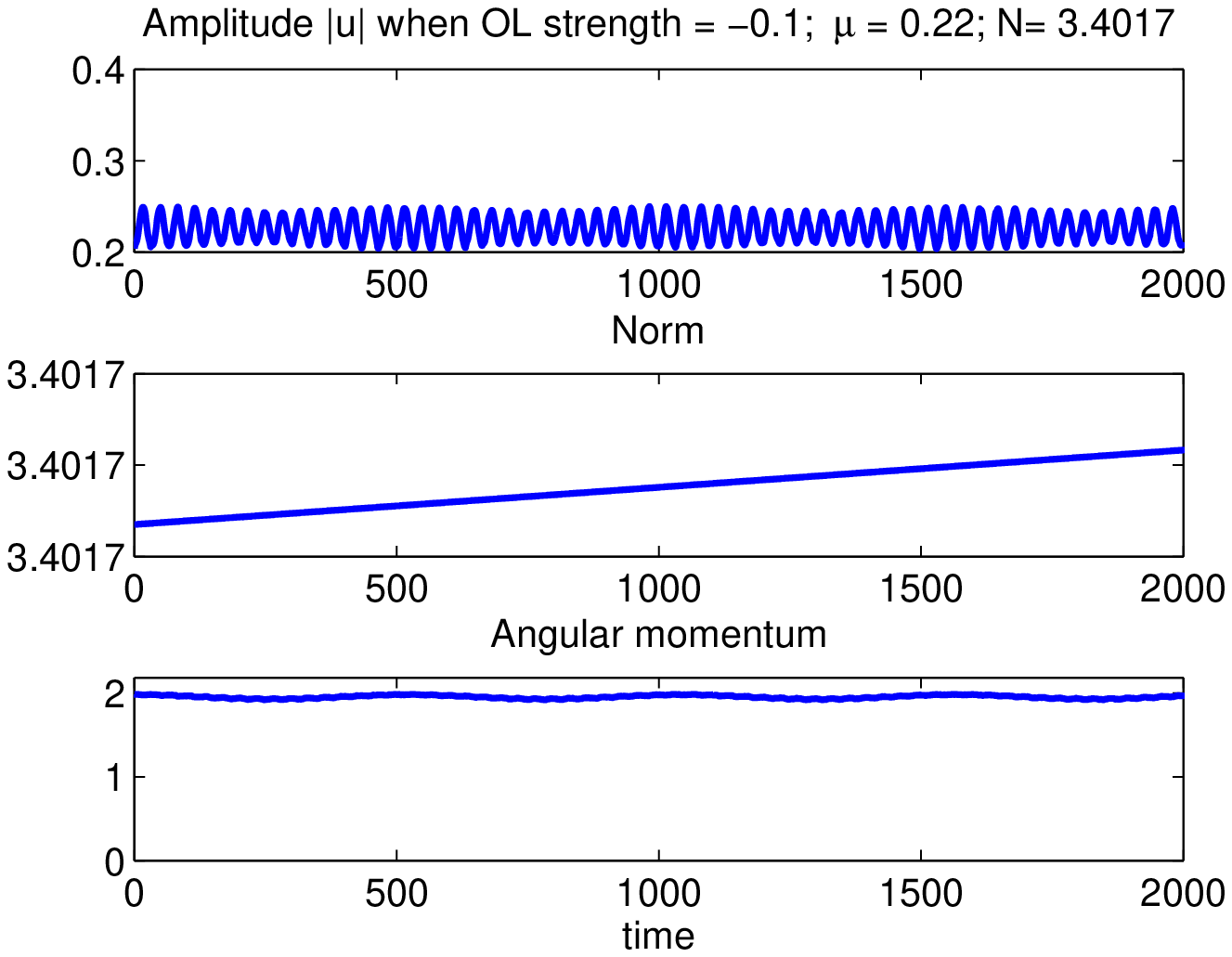, width=2.5in,height=2.5in}
   }
\caption{(Color online) The dynamics of a typical stable vortex soliton with
$S=1$ in the presence of a weak OL. Top: $\protect\varepsilon =0.1$, $%
N=4.6673$; bottom: $\protect\varepsilon =-0.1$, $N=3.4017$.}
\label{f:S1_weakOL1}
\end{figure}

The dynamics becomes much more complicated when the optical lattice is present.
For these cases, the main conclusions are as follows.

(i)The stability region becomes smaller for a weak OL ($%
\varepsilon =\pm 0.1$). Yet, the stable solutions still behave like a
coherent vortex ring, provided that norm $N$ is not too large. With an
increase of $N$, the solutions mimic the splitting-recombination scenario
observed in the absence of the OL (see above). Two typical solutions are
displayed in Fig.~\ref{f:S1_weakOL1}. 

\begin{figure}[tbp]
\centerline{
    \psfig{file = 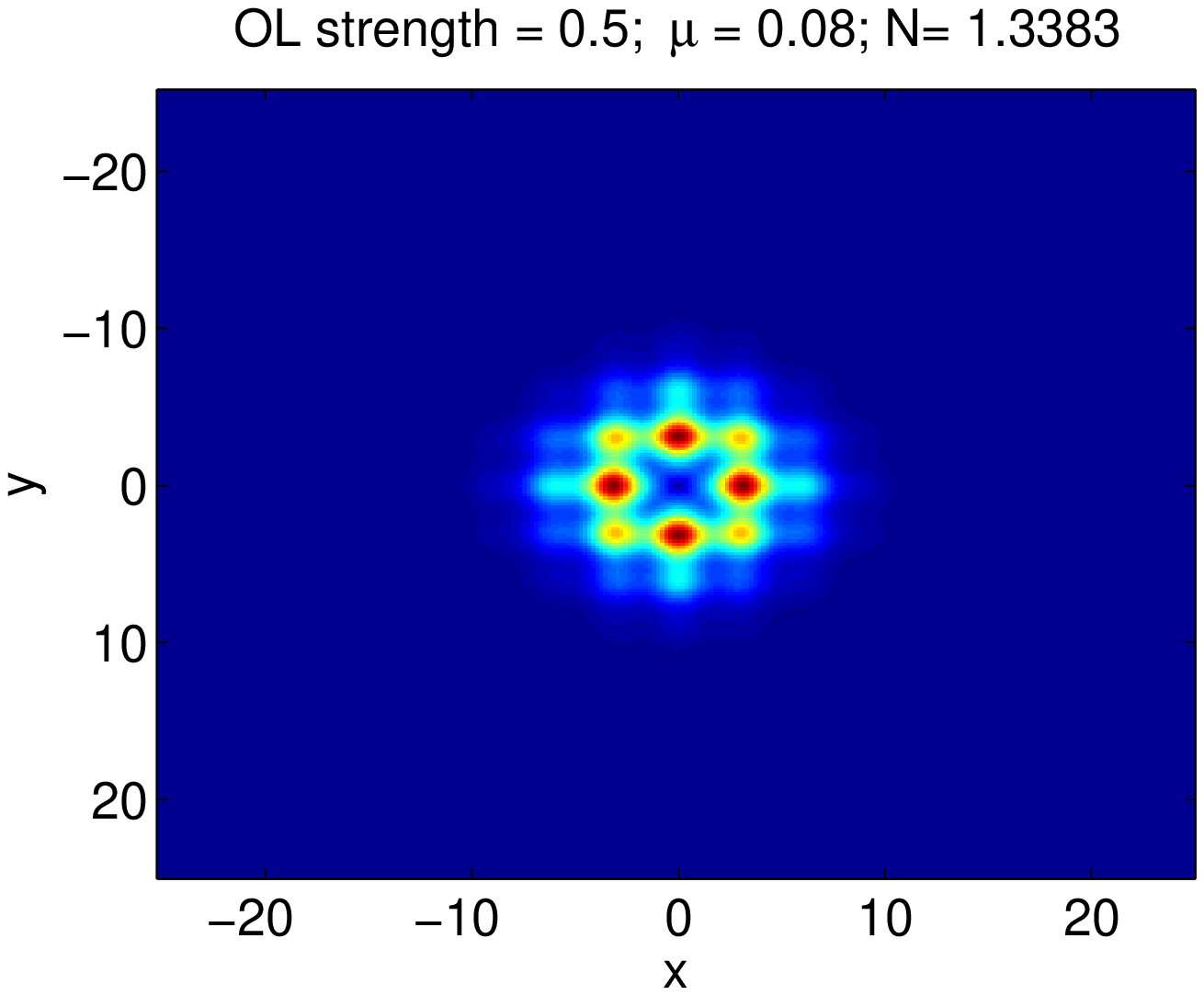, width=2.5in,height=2.5in}
\hspace{0.1in}
    \psfig{file = 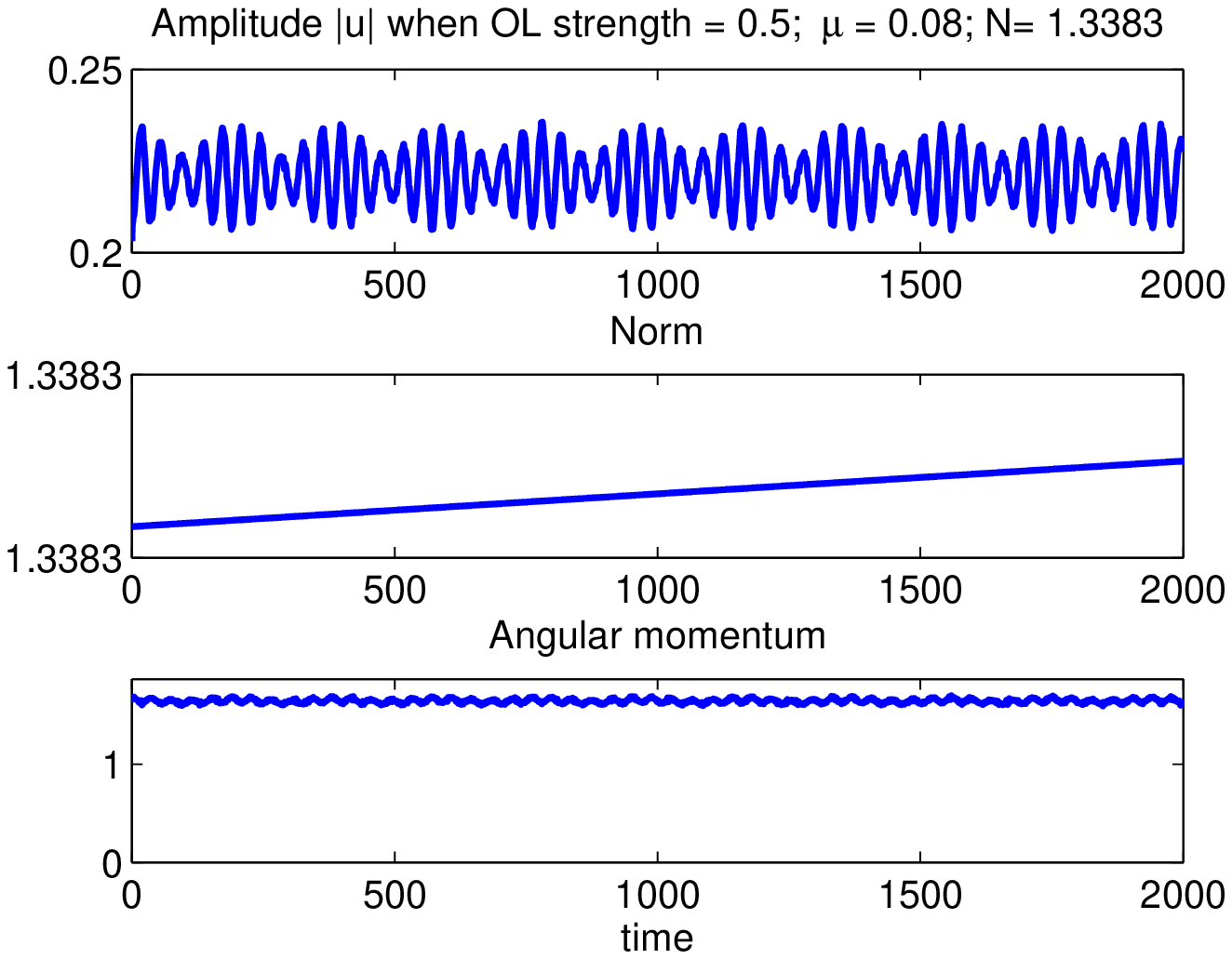, width=2.5in,height=2.5in}
   }
\centerline{
    \psfig{file = 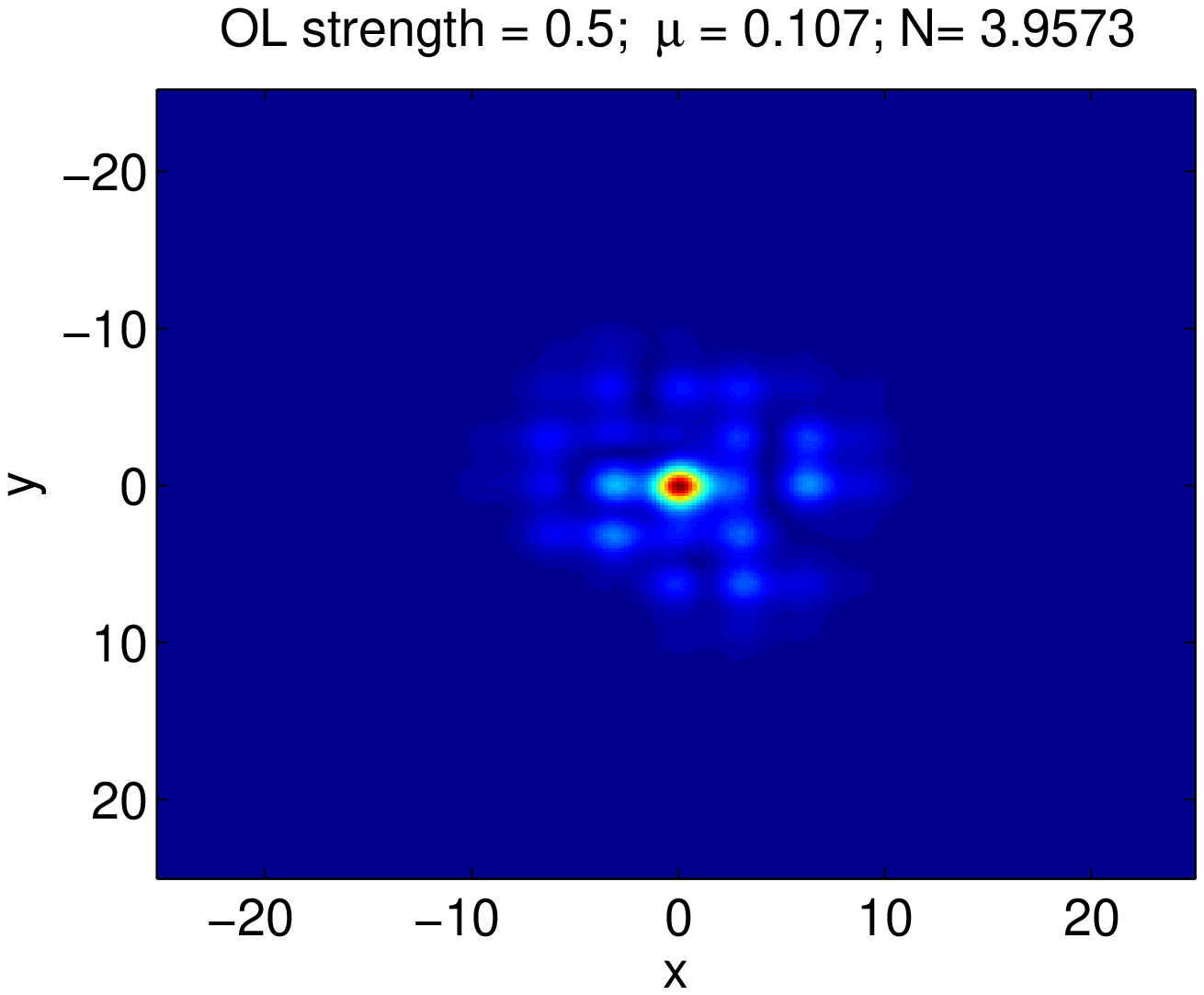, width=2.5in,height=2.5in}
\hspace{0.1in}
    \psfig{file = 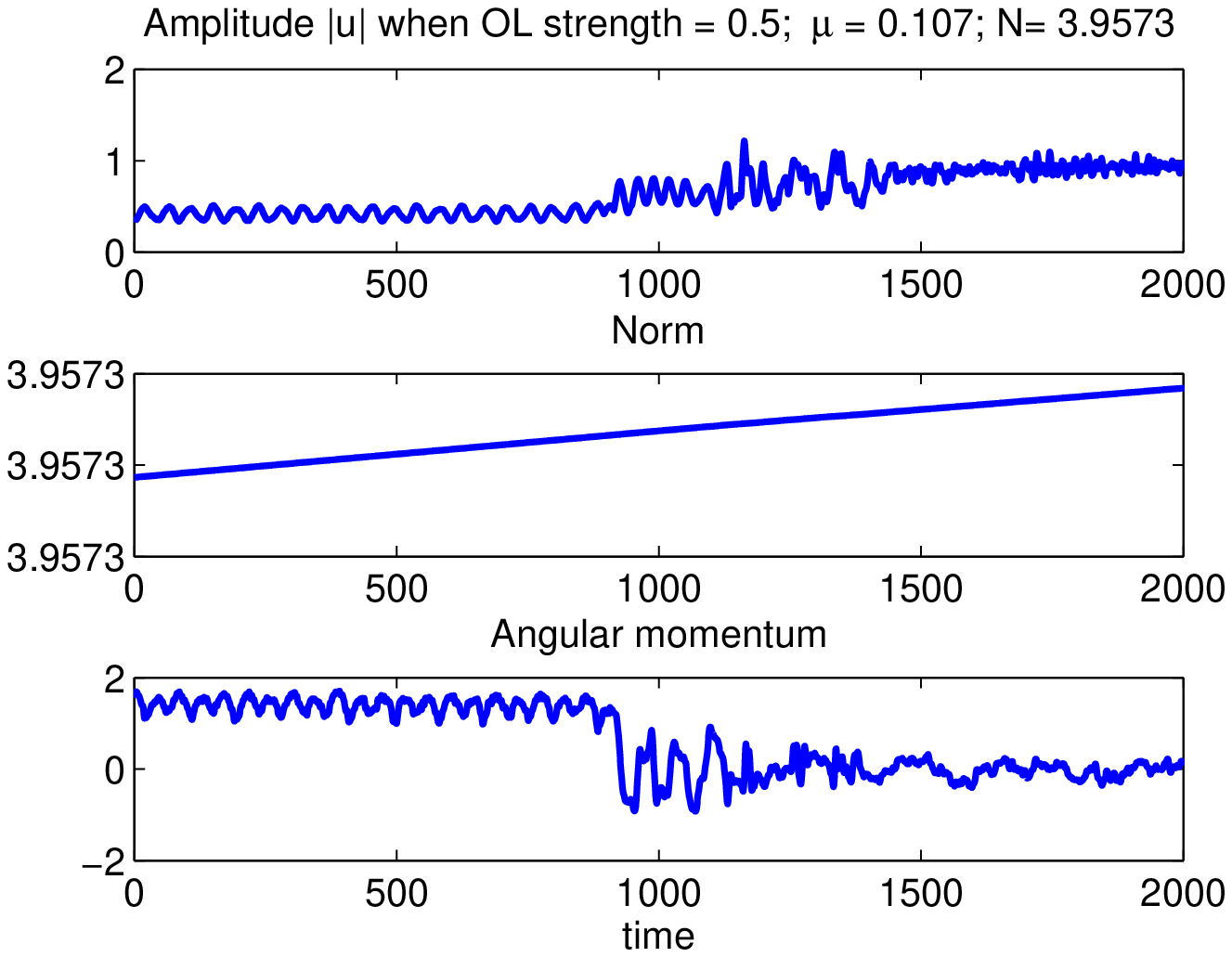, width=2.5in,height=2.5in}
   }
\caption{(Color online) A vortex soliton with topological charge
$S=1$, for the OL strength $\protect\varepsilon =0.5$. Top:
$N=1.3383$,  the final solution still being a vortex. Bottom:
$N=3.9573$, the solution evolving towards a fundamental soliton.}
\label{f:S1_OLE05}
\end{figure}

\begin{figure}[tbp]
\centerline{
    \psfig{file = 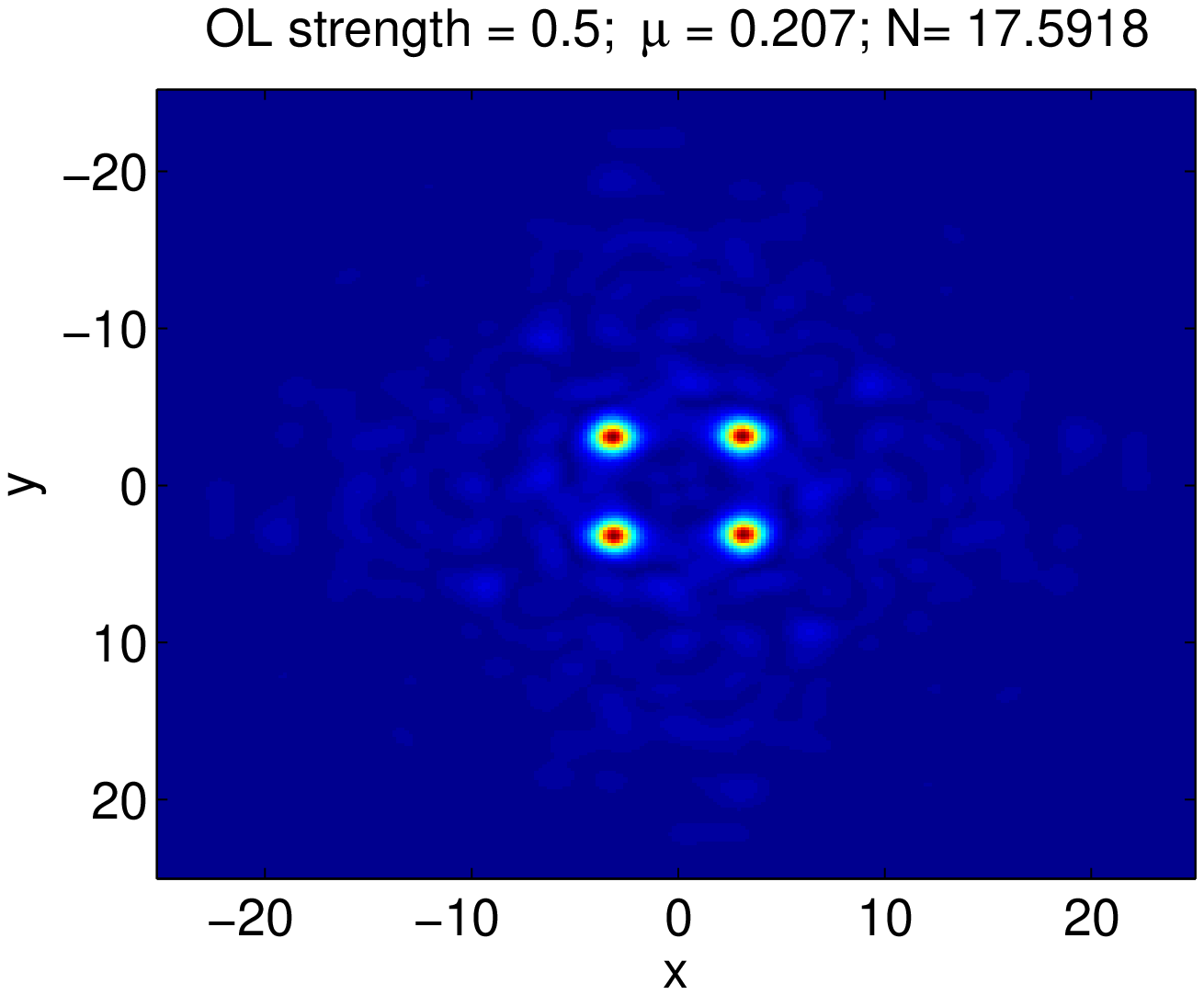, width=2.5in,height=2.5in}
\hspace{0.1in}
    \psfig{file = 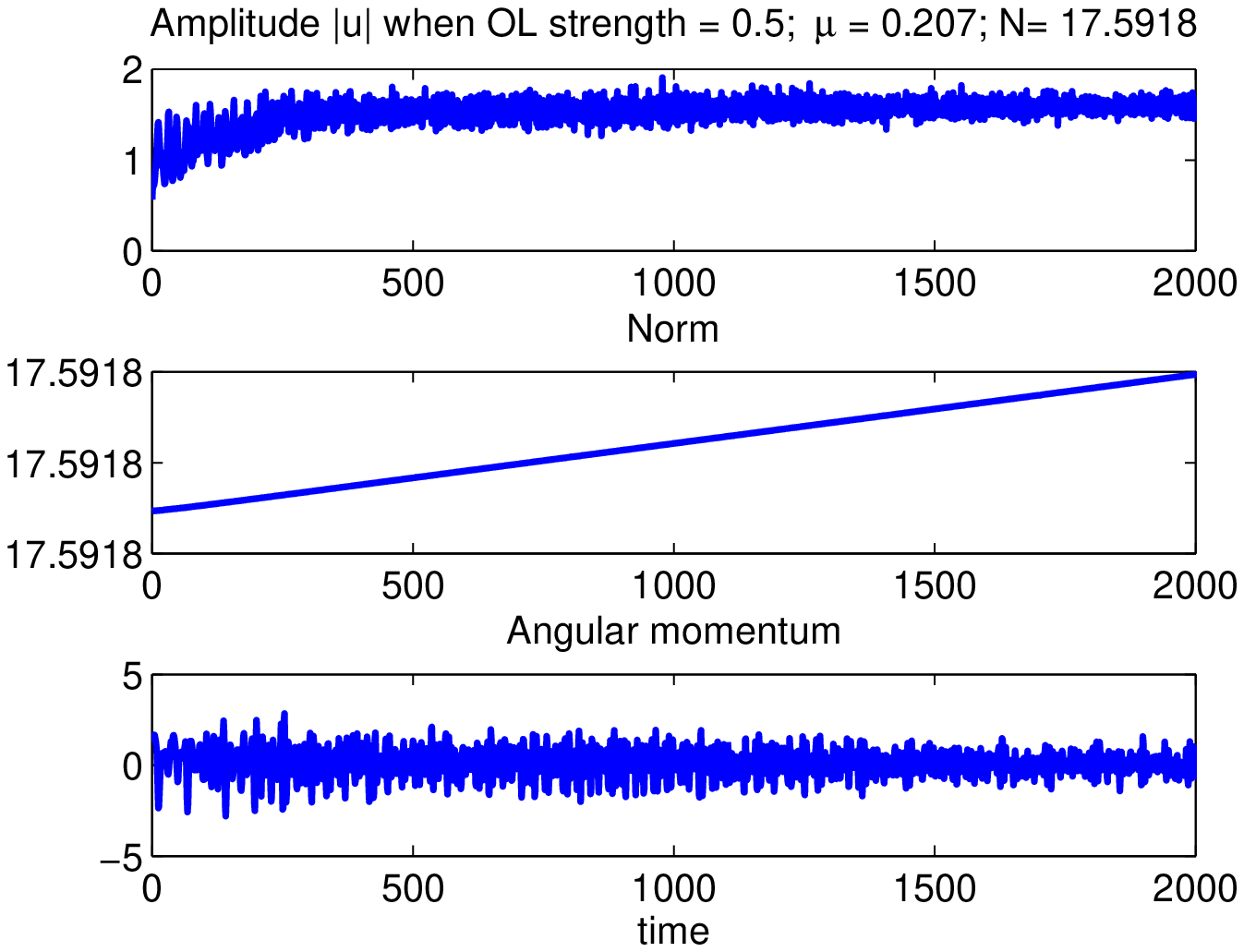, width=2.5in,height=2.5in}
   }
\caption{(Color online) A vortex solution from the stability region (\protect
\ref{2}). Here the lattice strength is $\protect\varepsilon =0.5$.}
\label{f:S1_OLE05_N17}
\end{figure}

\begin{figure}[tbp]
\centerline{
    \psfig{file = 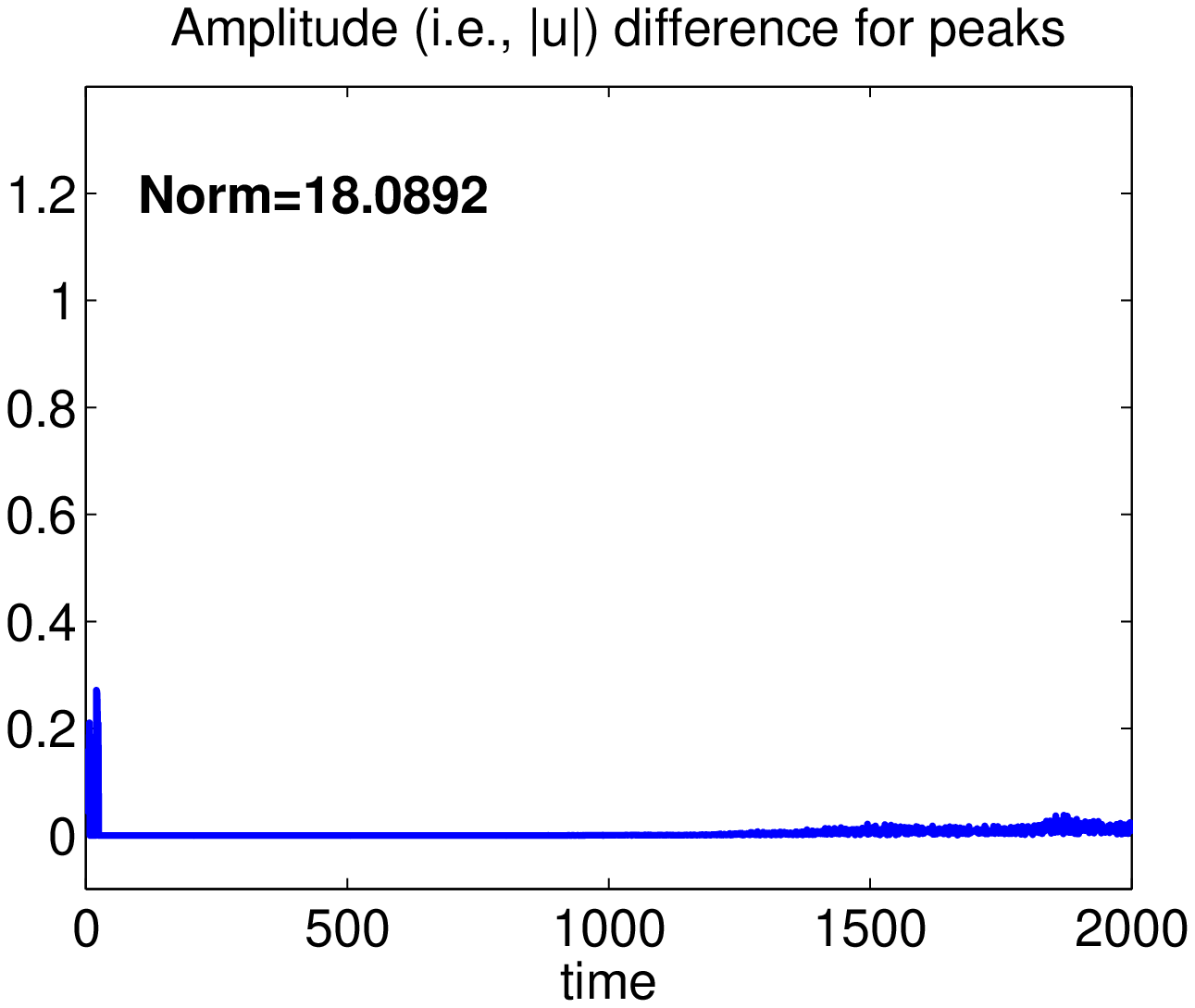, width=2.5in,height=2.5in}
\hspace{0.1in}
    \psfig{file = 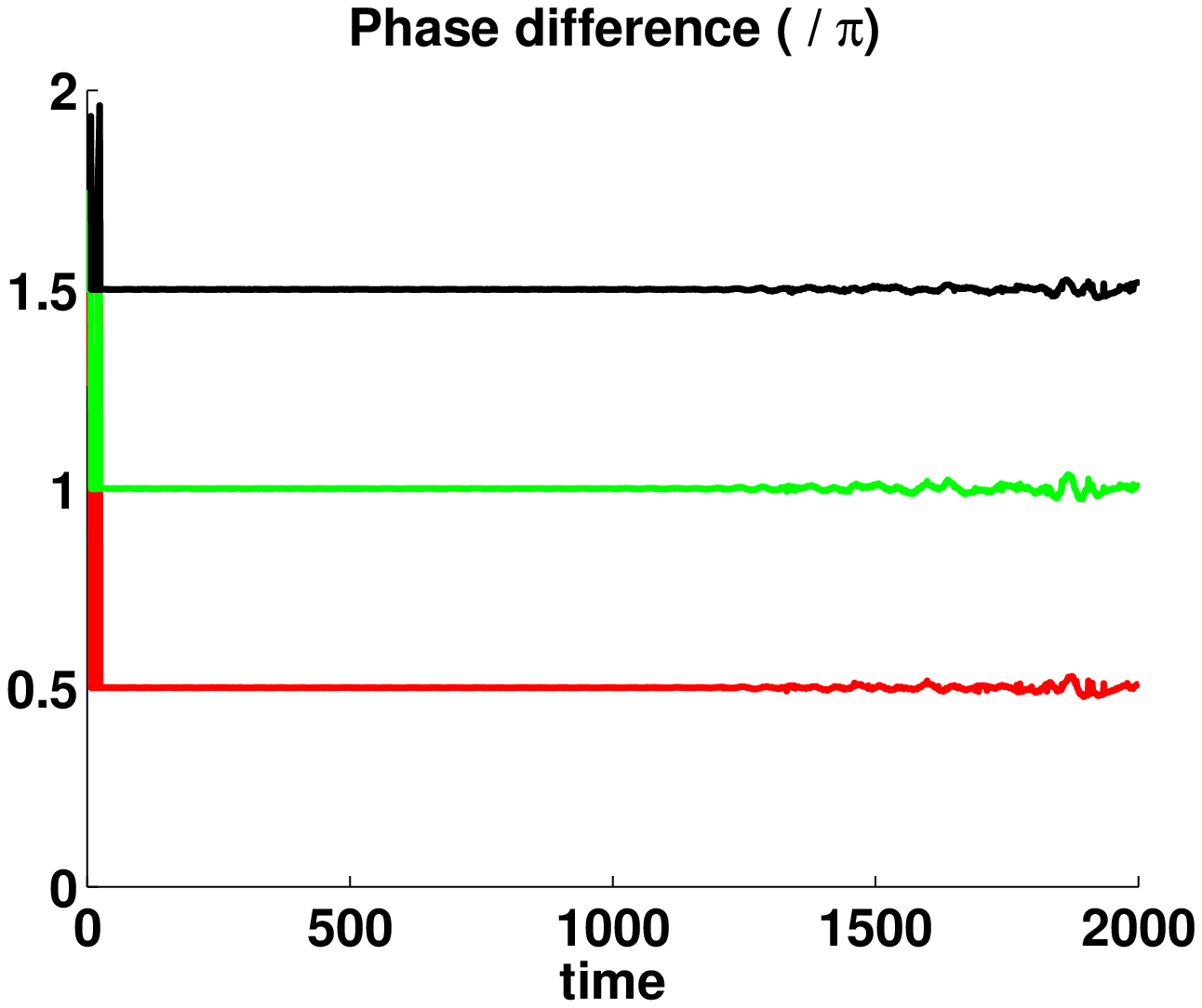, width=2.5in,height=2.5in}
   }
\centerline{
    \psfig{file = 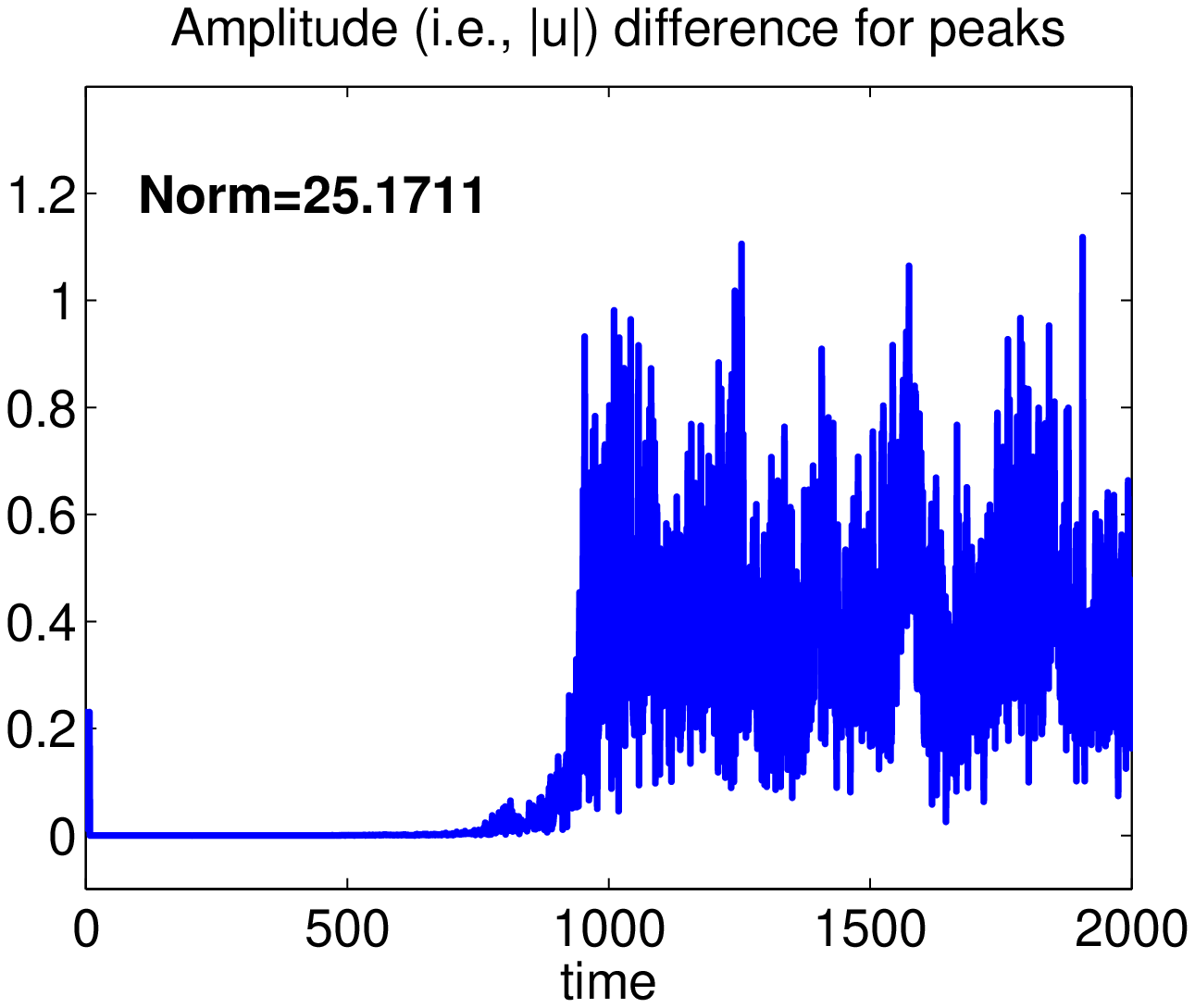, width=2.5in,height=2.5in}
\hspace{0.1in}
    \psfig{file = 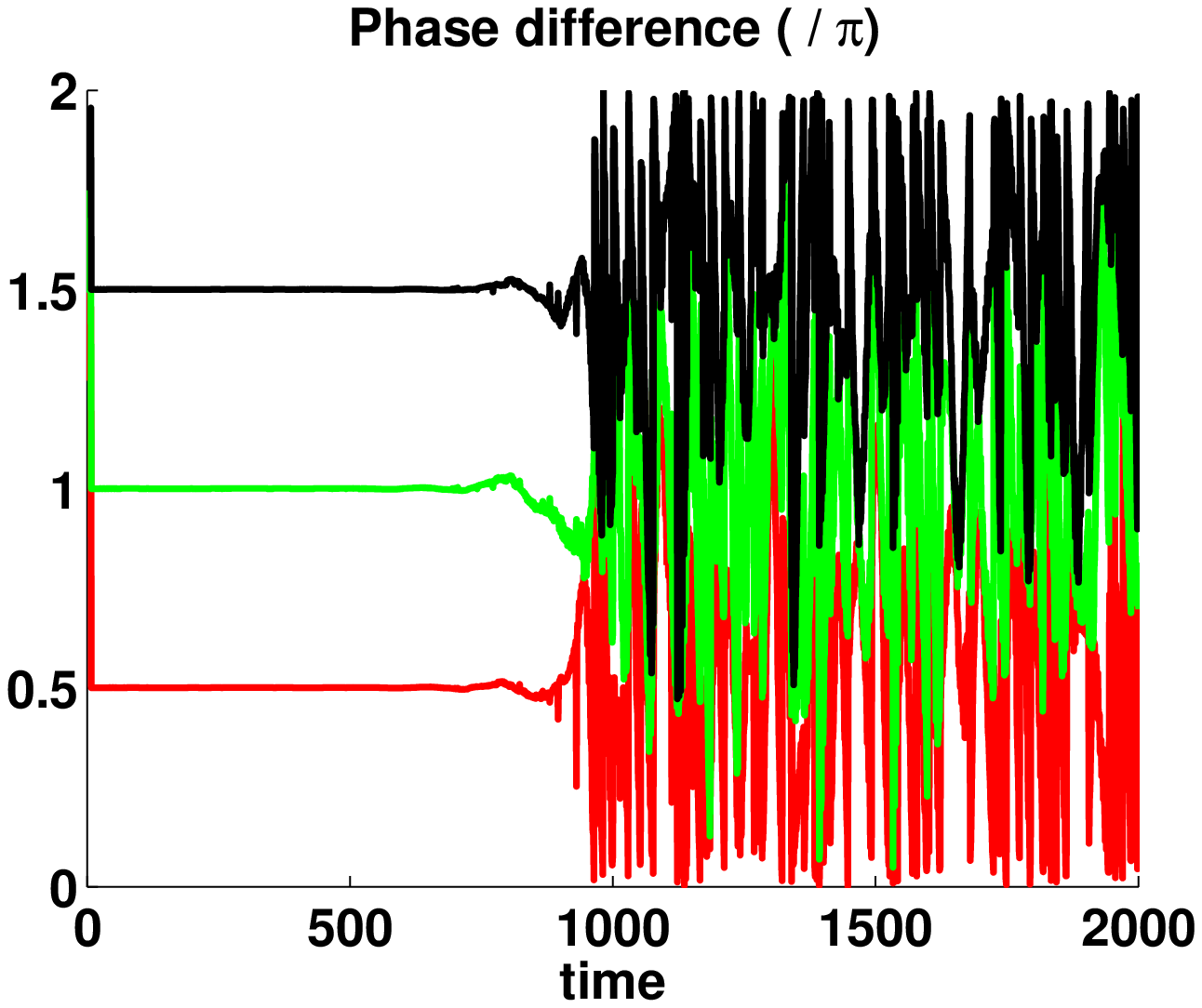, width=2.5in,height=2.5in}
   }
\caption{(Color online) Examples of the loss of symmetry for the
quasi-discrete vortices from the  stability region (\protect\ref{2}). The OL
strength is $\protect\varepsilon =0.5$. Left panels: the largest difference
among the four peak amplitudes. Right panels: phase shifts (in units of $%
\protect\pi $) between the peaks. Top: $N=17.92$; bottom: $N=25.17$.}
\label{f:symm}
\end{figure}

(ii) In the case of the OL with a moderate strength ($\varepsilon =\pm 0.5$%
), there are two regions in which the solution stays stable for up to $%
T=2000$. In particular, it is stable when its initial norm lies in two
intervals,
\begin{eqnarray}
0 &<&N\leq 8.04,  \label{1} \\
17.59 &\leq &N\leq 25.17.  \label{2}
\end{eqnarray}%
In the stability region (\ref{1}), there are two kinds of solutions observed.
When the norm is small enough,
all the solutions are similar to the one shown in the top panel of
Fig.~\ref{f:S1_OLE05}. Basically, it is built of eight peaks forming
two groups. The first group includes the peaks at the 3, 6, 9 and 12
o'clock positions, and this set is always present. The second group
consists of the remaining peaks set along the diagonals, which
breathe as a function of time (thus becoming more or less
noticeable). The overall topological charge is still preserved
\footnote{%
A video illustrating this case can be found at: \textbf{\text{%
http://www.math.umass.edu/\~{}qchen/S1\_OLE05\_N1.zip}\ }}. The
second kind of dynamics eventually leads to the loss of the
vorticity, and transformation of the vortex into a fundamental
soliton, as shown in the bottom panel of Fig.~\ref{f:S1_OLE05}. Its
location may vary, depending on the initial norm of the solution.
The largest observed norm contained in such a stable peak is
$N=5.43$. When $N$ is increased to be out of the stability region
(\ref{1}), the solution eventually collapses. In the stability
region (\ref{2}), the solutions for all the considered
configurations feature four peaks. The largest total norm is
$N=21.53$, with
$5.38$ in each individual peak. A solution of this type is displayed at Fig.~%
\ref{f:S1_OLE05_N17}. However, only some of the stable solutions are true
vortices (marked by circles in Fig.~\ref{f:SR_S1}), bearing the
topological charge of $S=1$. For other solutions, the loss of the global
coherence among the peaks occurs in the course of the evolution, at the same
moment of time when the symmetry-breaking occurs, i.e., the peaks start to
have different height.
Two examples are presented in Fig.~\ref{f:symm}, with one preserving the vorticity (similar to the vortices on 
discrete lattice, although there is complete lack of the
isotropy in the system \cite{2001,pgk_book}), and one becoming incoherent
and thus shedding the vorticity off.

(iii) For large strength of the OL (e.g., $\varepsilon =\pm 0.9$), the first
stability region, corresponding to Eq. (\ref{1}), expands, similar to what
is the case for the fundamental states in Section III. On the other hand,
the second stability region, which corresponds to Eq. (\ref{2}), practically
disappears. The final stable solutions (at $T=2000$) exhibit breathing
behavior. Yet aside from their more pronounced peaks, they do not exhibit
any salient structural differences from their counterparts considered above
at intermediate values of $\varepsilon $.

\section{Conclusions and future challenges}

In this work, we have examined the quenched dynamics of both the fundamental
states and vortices with topological charge $S=1$ in BEC. The quench 
consists of the
sudden reversal of the nonlinearity sign from repulsive to attractive. The
resulting states were investigated by systematic simulations, addressing
both the impact of parameters, such as the strength of the OL (optical
lattice), and the effect of initial conditions, by considering a range of
values of the initial norm, $N$ (which is proportional to the number of
atoms in the BEC). A principal result is that the OL expands the range of
initial norms which do not lead to the collapse of both the fundamental and
vortex states. In fact, this expansion occurs well over the interval of norms
for which static vortices were\ previously identified as stable states via
the linear stability analysis. For the fundamental states, the application
of the VA (variational approximation) is more efficient for the lower norm
of the initial state. Additional beating effects, not captured by the VA,
were found close to the collapse threshold. On the other hand, a particular
finding, in the case of the vortex, is that, in addition to the regimes
where a vortex survives in the breathing form or collapses, there is an
intermediate regime where the vortex (in the presence of the OL) loses its
topological character, yet remains immune to the collapse. Furthermore, a
second stability region was identified for the vortex, in the presence of
the OL with an intermediate strength, where the collapse was avoided due to
the formation of a robust quasi-discrete vortex.

The same system may also be implemented in nonlinear optics, where the
combination of the HO trap and OL potential corresponds to photonic-crystal
fibers, while the switch of the nonlinearity corresponds to a junction of
two waveguides made of self-defocusing and self-focusing materials \cite%
{Isaac}.

We believe that these results provide a potential for further studies on
this theme, both at the theoretical and at the experimental level. At the
theoretical level, it may be useful to develop 3D generalizations of the
analysis developed in this paper, either for the full case of spherically symmetric
BECs or for strongly anisotropic traps. The effect of anisotropy in 2D
would be interesting to examine too, as it departs from
the configuration of the isotropic cylindrical trap. On the other hand, at
the experimental level, recent advances in producing~\cite{BPA_KZ,BPA_10} and
monitoring \cite{freilich,ourvort3} the dynamics of vortices and vortex
clusters, in conjunction with the well-established control of the
BEC dynamics by means the Feshbach resonance~\cite{randyh}, render
particularly appealing and accessible examination of such quenches (or
the corresponding adiabatic transitions) between the repulsive and
attractive regimes.

\textbf{Acknowledgments}. The work of Q.Y.C. was supported, in a part, by
the National Science Foundation under the grant DMS-1016047. P.G.K. and
B.A.M. appreciate a support provided by the Binational (US-Israel) Science
Foundation through grant No. 2010239.

        \bibliographystyle{abbrv}

        \bibliography{cqy}

\end{document}